\documentclass[a4paper,11pt]{article}
\usepackage[utf8]{inputenc}
\usepackage{geometry}
\usepackage{textcomp}
\usepackage{amsmath}
\usepackage{amssymb}
\usepackage{graphicx}
\usepackage{color,xcolor}
\usepackage{mathrsfs}
\usepackage{caption}
\usepackage[colorlinks=true, allcolors=blue]{hyperref}
\usepackage{tcolorbox}
\usepackage{tikz}
\usepackage{tikz-cd}
\usepackage{physics}
\usepackage{diagbox}
\usepackage{listings}
\usepackage{array}
\usepackage{color}
\usepackage{subfig}
\newcommand{\nobracket}{}
\newcommand{\tmop}[1]{\ensuremath{\operatorname{#1}}}

\usepackage{cite}

\usepackage{cancel}
\usepackage{multirow}
\usepackage{extarrows}
\allowdisplaybreaks[2]
\numberwithin{equation}{section}
\usepackage[normalem]{ulem}
\usepackage{color}
\definecolor{darkgreen}{RGB}{40,150,60}

\begin{document}
\title{ Path-integral quantization of  tensionless (super) string}
\vspace{14mm}
\author{
Bin Chen$^{1,2,3}$, Zezhou Hu$^1$, Zhe-fei Yu$^3$ and Yu-fan Zheng$^1$\footnote{bchen01@pku.edu.cn, z.z.hu@pku.edu.cn, yuzhefei@pku.edu.cn, 1801110091@pku.edu.cn}
}
\date{}

\maketitle

\begin{center}
	{\it
		$^{1}$School of Physics and State Key Laboratory of Nuclear Physics and Technology,\\Peking University, No.5 Yiheyuan Rd, Beijing 100871, P.~R.~China\\
		\vspace{2mm}
		$^{2}$Collaborative Innovation Center of Quantum Matter, No.5 Yiheyuan Rd, Beijing 100871, P.~R.~China\\
		$^{3}$Center for High Energy Physics, Peking University, No.5 Yiheyuan Rd, Beijing 100871, P.~R.~China
	}
	\vspace{10mm}
\end{center}

\begin{abstract}
In this work, we study the  tensionless (super)string in the formalism of path-integral quantization. We introduce BMS $bc$ and $\beta\gamma$ ghosts intrinsically by accounting for the Faddeev-Popov determinants appeared in fixing the gauges. We then do canonical quantization and obtain the critical dimensions for different tensionless strings. We find that among four kinds of tensionless superstrings, the $\mathcal{N}=2$ homogeneous and inhomogeneous doublet tensionless superstrings have the same critical dimension as the usual superstrings. Taking the  BMS $bc$ and $\beta\gamma$ ghosts as new types of BMS free field theories, we find that their enhanced underlying symmetries are generated by  BMS-Kac-Moody algebras, with the Kac-Moody subalgebras being built from a three-dimensional
non-abelian and non-semi-simple Lie algebra.

\end{abstract}

\baselineskip 18pt
\newpage

\tableofcontents{}

\newpage

\section{Introduction}
It is well-known that the  string scattering amplitudes behave in a particularly simple way in the high energy limit $\alpha'\to \infty$ \cite{Gross:1987ar,Gross:1987kza}.  There are infinitely
many linear relations among these scattering amplitudes, which indicate the existence of a
higher symmetry structure in this tensionless limit \cite{Gross:1988ue}.  In a modern perspective, the study of tensionless strings had been linked with the emergence of higher-spin symmetries \cite{Sundborg:2000wp}.  The spectrum of tensile string theory includes an infinite tower of massive
particles of arbitrary spin. In the tensionless limit, all these particles become massless and the theory is thought to exhibit higher-spin symmetry. In fact,  the tensionless limit of string theory has been related to Vasiliev's higher spin theory \cite{Vasiliev:2003cph}, leading to  several proposals of holographic dualities \cite{Klebanov:2002ja, Sezgin:2002rt, Gaberdiel:2010pz,Chang:2012kt,Gaberdiel:2014cha}. More recently, an exact AdS$_3$/CFT$_2$ correspondence has been proposed \cite{Eberhardt:2018ouy,Eberhardt:2019ywk}, where the authors gave strong 
evidences that  the tensionless limit of type \uppercase\expandafter{\romannumeral2}B string theory on AdS$_3\times $S$^3\times $T$^4$ is dual to the symmetric orbifold CFT Sym$^N$(T$^4$). 

While the study of tensionless string in AdS is very interesting and important, especially for the AdS/CFT correspondence, the tensionless (super)string in flat spacetime, which is often called null strings since the early work of Schild \cite{Schild:1976vq}, has drawn much attention in the past few years. In fact, there have been many early efforts towards the quantization of the tensionless (super)string \cite{Karlhede:1986wb,Lizzi:1986nv,Amorim:1987bk,Gamboa:1989px,Gamboa:1989zc,Gustafsson:1994kr,Lindstrom:2003mg}. These works started from Schild's action of the tensionless string \cite{Schild:1976vq}. In contrast, the recent studies on the tensionless string are based on the ILST action \cite{Isberg:1993av} (or its supersymmetric generalization) of the null string. The advantage of the ILST action is that  two-dimensional(2D) Galilean conformal symmetry or equivalently the Bondi-Metzner-Sachs (BMS) symmetry arises as the residual gauge symmetry on the tensionless worldsheet \cite{Bagchi:2013bga}. Since 2D conformal symmetry 
offers a guiding principle for the construction of the  usual tensile strings in the conformal gauge \cite{Friedan:1985ge}, one  expects that BMS symmetry will be important for the  tensionless string in the same spirit.  Moreover, since the BMS symmetry arise as the asymptotic symmetry group of flat spacetime, the tensionless (super)string could be related to the flat-space holography. Besides, the null string turns out to be closely related to the ambitwistor string \cite{Casali:2016atr, Casali:2017zkz} (see also \cite{Siegel:2015axg}).

As mentioned above, the tensionless string is closely related to the BMS field theories (BMSFT). The  BMS field theories is a type of non-relativistic conformal field theories  with the following scaling symmetry and 
boost symmetry, 
\begin{equation}
\begin{aligned}
    x\rightarrow \lambda x, &\qquad y\rightarrow \lambda y,\\
x\rightarrow x,& \qquad y\rightarrow y+v x.
\end{aligned}
\end{equation}
The symmetry algebra can be enhanced to the two dimensional Galilean conformal algebra \cite{Chen:2019hbj}, which is isomorphic to the BMS algebra in three dimensions.  The generators of  GCA$_2$ (BMS$_3$) include the superrotations $L_n$ and the supertranslations $M_n$, satisfying the following commutation relations
\begin{equation}
\begin{aligned}
&&[L_n,L_m]&=(n-m)L_{n+m}+\frac{c_L}{12} n(n^2-1)\delta_{n+m,0},\\
&&[L_n,M_m]&=(n-m)M_{n+m}+\frac{c_M}{12} n(n^2-1)\delta_{n+m,0},\\
&&[M_n,M_m]&=0.
\end{aligned}
\end{equation}
The algebra is of infinite dimensions, just like the Virasoro algebra in CFT$_2$. The BMS$_3$/GCA$_2$ isomorphism motivates a lot of works establishing the holography in asymptotic flat spacetimes, see e.g. \cite{Bagchi:2012cy,Barnich:2012xq,Bagchi:2012xr,Bagchi:2014iea,Jiang:2017ecm,Hijano:2018nhq,Hijano:2019qmi,Apolo:2020bld,Apolo:2020qjm}.
On its own right, BMSFT has been widely studied in the  past few years. These theories are typically not unitary and exhibit novel features \cite{Bagchi:2009pe,Chen:2020vvn,Chen:2022jhx,Saha:2022gjw}, especially the appearance of the boost multiplet \cite{Chen:2020vvn}. Nevertheless, the BMS (GCA) bootstrap seems still viable \cite{Chen:2020vvn,Chen:2022cpx,Chen:2022jhx,Bagchi:2016geg,Bagchi:2017cpu}. Very recently, a few concrete models of BMSFT had been constructed \cite{Hao:2021urq,Yu:2022bcp,Hao:2022xhq,Banerjee:2022ocj}, which  are in fact closely related to the tensionless string\footnote{Besides these theories, early examples of BMSFT called the generalized Galilean free field theories were constructed in \cite{Chen:2020vvn}. These theories only exhibit global BMS symmetry, similar to the generalized free field theories in CFT.}.

In this work, we would like to study the tensionless bosonic string and superstring from the path integral point of view. There have been some studies on the tensionless bosonic and superstring from null string point of view \cite{Bagchi:2015nca,Bagchi:2016yyf,Bagchi:2017cte,Bagchi:2018wsn,Bagchi:2020fpr,Bagchi:2021rfw,Bagchi:2022iqb} since the discovery of BMS symmetry in the ILST action in \cite{Bagchi:2013bga}. In particular, the quantum theory of the tensionless bosonic string had been studied in \cite{Bagchi:2020fpr} and three  distinct choices of tensionless vacua had been found (see also the early work \cite{Hwang:1998gs} for a systematic discussion of  different choices of vacua). Then the lightcone quantisation had been performed  for these theories to find the critical dimensions \cite{Bagchi:2021rfw}. In the present work, we start with doing the path integral quantization of the tensionless bosonic strings and reproduce the results of critical dimensions in \cite{Bagchi:2021rfw}\footnote{More precisely, only the results of the induced vacuum and the flipped vacuum can be reproduced. The critical dimension of the oscillating vacuum is not accessible by the cancellation of the BMS conformal anomaly.}. When doing the path integral, we introduce the  BMS $bc$-ghost, which turns out to be the inhomogenous ultra-relativistic(UR) limit of the usual $bc$-ghost. This theory is another BMS free theory apart from the BMS free scalar \cite{Hao:2021urq} and fermion \cite{Yu:2022bcp,Hao:2022xhq,Banerjee:2022ocj}.  We also find another BMS $bc$-ghost by taking the homogenious UR limit of the usual $bc$-ghost. However, this theory is not related to the  path integral of the tensionless bosonic string and only contains boost singlet representations. 

Furthermore, we discuss the path-integral quantization of the tensionless superstring. The tensionless superstring had been studied in \cite{Bagchi:2016yyf,Bagchi:2017cte,Bagchi:2018wsn}, where two versions of the BMS fermion action were proposed, leading to homogenous and inhomogenous tensionless superstrings respectively. We give a more complete classification of the quantum tensionless superstrings by including different possible amounts of supersymmetries as well as  different kinds of gravitino fields. We introduce BMS $\beta\gamma$ ghosts into the study and do canonical quantization on different tensionless superstrings, and calculate their critical dimensions by imposing the conformal anomaly cancellation condition.  We find that not all of these theories are consistent because the critical dimensions in some cases are not integers. The non-integer critical dimensions do not appear in the usual tensile superstrings, so these  tensionless superstrings with non-integer critical dimensions can not be obtained by taking the tensionless limit of the tensile superstrings.  Moreover, the BMS $bc$ and $\beta\gamma$ ghost field theories present novel examples of  BMS free fields and deserve more careful studies. We show that there exist underlying enhanced symmetries in these theories, which are generated by  BMS-Kac-Moody algebras. In particular, they include an anisotropic scaling symmetry. This is similar to the BMS free scalar and fermion \cite{Yu:2022bcp}. 

It would be illuminating to compare our results with the ambitwistor (super)strings. Classically, the ambitwistor string is a gauge-fixed version of the null string \cite{Casali:2016atr}. At the quantum level, this equivalence holds when quantizing the null string in the flipped vacuum or  ``normal-ordering prescription'' (terminology used in  \cite{Casali:2016atr}).  Relations between tensionless superstrings and  ambitwistor strings had also been discussed in \cite{Kalyanapuram:2021vjt, Kalyanapuram:2021xow}, at the level of  chiral superstring integrands. The  $\mathcal{N}=2$ ambitwistor  superstring had been discussed by Mason and Skinner \cite{Mason:2013sva}\footnote{In \cite{Mason:2013sva}, they also  discussed the bosonic and $\mathcal{N}=1$ cases.}, and its critical dimension was found to be $d=10$. For the ambitwistor strings with different possible amounts of supersymmetries\cite{Ohmori:2015sha}, their critical dimensions in fact coincide with our results of homogeneous tensionless superstrings (see Table \ref{tb:CriticalDimForHomogeneous})\footnote{According to the result $c^m=52-11N$ below Eq. (3.5) in \cite{Ohmori:2015sha}, it is easy to find exactly the same critical dimensions as in Table \ref{tb:CriticalDimForHomogeneous}.}. This coincidence reflects the equivalence between the ambitwistor (super)strings and the null (super)strings \cite{Casali:2016atr}. Besides, our study includes   more types of   tensionless superstrings, whose critical dimensions are given in Table \ref{tb:CriticalDimForInhomoDoublet} and Table \ref{tb:CriticalDimForInhomoSinglet}.

\subsection*{Outline of the paper}
The remaining parts of this paper are organized as follows. In the next section, we discuss the path-integral quantization of tensionless bosonic string and introduce BMS $bc$ ghosts. In section \ref{section3}, we turn to the study of the path-integral of tensionless superstring. We construct the actions of BMS $\beta\gamma$ ghosts for various tensionless superstrings, and compute the critical dimensions  using the canonical quantization. In section \ref{multiplet}, we investigate the underlying symmetry algebra of BMS $bc$ and $\beta\gamma$ ghosts. We end with some discussions in section \ref{section5}.  In the appendix, we use the lightcone quantization to compute the critical dimension of the inhomogeneous doublet tensionless superstring, as a consistent check. 

\section{Tensionless bosonic string and BMS \texorpdfstring{$bc$}~ ghosts}

In this section, we discuss the  path-integral quantization of the tensionless bosonic string. Similar to the usual tensile string, we need to introduce the BMS version of  $bc$-ghost for the  Faddeev-Popov determinant.  We will show that this intrinsic BMS  $bc$-ghost can be obtained by taking the inhomogeneous ultra-relativistic limit of the usual  $bc$-ghost.  By contrary, we can obtain  the homogeneous BMS $bc$-ghosts by taking the homogeneous UR limit as well, but find that they have nothing to do with the path-integral of the tensionless bosonic string.

\subsection{Path-integral and the BMS \texorpdfstring{$bc$}~-ghost}

We start with the ILST action \cite{Isberg:1993av},
\begin{equation}
  S(V,X)=\frac{1}{2 \pi}\int d^2 \sigma ( V^{\alpha} V^{\beta} \partial_{\alpha} X
  \cdot \partial_{\beta} X),
\end{equation}
Here $V$ is a vector density of weight $-\frac{1}{2}$ describing the geometry as $\sqrt{- g} g^{a b} = V^a V^b$. Under  an infinitesimal diffeomorphism $\sigma^\alpha\to \sigma^\alpha+\epsilon^\alpha$,  it transforms as
\begin{equation}
\delta_\epsilon V^\alpha=-V^\gamma \partial_\gamma \epsilon^\alpha+\epsilon^\gamma \partial_\gamma V^\alpha+\frac{1}{2} V^\alpha \partial_\gamma \epsilon^\gamma.
\end{equation}
 The partition function is 
\begin{equation}
\begin{aligned}
  Z&=\int D\xi D V D X e^{i S(V,X)},
\end{aligned}
\end{equation}
where  $\xi$ represents the BMS transformation keeping $V$ invariant, $V^{\xi} \equiv V+\delta_\xi V= V$, and it is given by
\begin{equation}
\begin{aligned}
  \xi^0 & = f' (\sigma) \tau + g (\sigma),\\
  \xi^1 & = f (\sigma).
\end{aligned}
\end{equation}
In the partition function, $D\xi$ is the integration over different geometries of the worldsheet and $D V$ is the integration over the degrees of freedom in the diffeomorphism. 

We may transform the integration $D V$  into the integration of an intrinsic diffeomorphism parameter $\epsilon$ by using the following identity 
\begin{equation}\label{identtty}
  1=\int D \epsilon \delta \left( V^{\epsilon} - V^{0}
  \right) \det \left( \frac{\delta V^{\epsilon}}{\delta \epsilon} \right),
\end{equation}
where $V^{\epsilon}$ is the diffeomorphism transformation of $V^0$ and we have chosen the gauge to set  $V^{0} = (1,0)$. Inserting Eq.(\ref{identtty}) into the  partition function, we obtain
\begin{equation}
\begin{aligned}
Z&=\int D \epsilon D \xi D V D X \delta (V^{\epsilon} -
  V^{\tmop{gauge}}) \det \left( \frac{\delta V^{\epsilon}}{\delta \epsilon}
  \right) e^{- S (V, X)}\\
  &=\int D \epsilon D \xi D V D X \delta (V^{\epsilon} -
  V^{\tmop{gauge}}) \det \left( \frac{\delta V^{\epsilon}}{\delta \epsilon}
  \right) e^{- S (V^{\epsilon}, X^{\epsilon})}.
\end{aligned}
\end{equation}
If there is no anomaly for diffeomorphism, i.e. $D V D X=D V^\epsilon D X^\epsilon$, it turns out that
\begin{equation}
\begin{aligned}
Z&=\int D \epsilon D \xi D V^{\epsilon} D X^{\epsilon} \delta
  (V^{\epsilon} - V^{\tmop{gauge}}) \det \left( \frac{\delta
  V^{\epsilon}}{\delta \epsilon} \right) e^{- S (V^{\epsilon},
  X^{\epsilon})}\\
  &=\int D \epsilon \int D \xi D V D X \delta (V - V^{\tmop{gauge}}) \det
  \left( \frac{\delta V}{\delta \epsilon} \right) e^{- S (V, X)}\\
  Z &\rightarrow Z / \int D \epsilon\\
  & =\int D \xi D V D X \delta (V - V^{\tmop{gauge}}) \det \left(
  \frac{\delta V}{\delta \epsilon} \right) e^{- S (V, X)}\\
  & =\int D \xi D X \det \left( \frac{\delta V}{\delta \epsilon} \right)
  e^{- S (V, X)}.
\end{aligned}
\end{equation}
In the last step $V = V^{\tmop{gauge}}$ has been imposed. If there is no anomaly
for Weyl (BMS) transformation parameterized by $\xi$, then
\begin{equation}
  Z\rightarrow \sum_{\tmop{topology}} \int D X \det \left( \frac{\delta
  V}{\delta \epsilon} \right) e^{- S (V, X)}.
\end{equation}
For simplicity, we focus on the case that the worldsheet is a null-cylinder, and have 
\begin{equation}
\begin{aligned}
  Z&=\int D X \det \left( \frac{\delta V}{\delta \epsilon} \right) e^{- S
  (V, X)}=\int D X \det (\Delta^{\beta}_{\alpha}) e^{- S (V, X)},
\end{aligned}
\end{equation}
where
\begin{equation}
  \Delta^{\beta}_{\alpha} \equiv \frac{\delta V^{\beta}}{\delta
  \epsilon^{\alpha}} = - \delta^{\beta}_{\quad \alpha} V^{\gamma}
  \partial_{\gamma} + \partial_{\alpha} V^{\beta} + \frac{1}{2} V^{\beta}
  \partial_{\alpha} .
\end{equation}

\subsubsection{BMS \texorpdfstring{$bc$}~-ghost action and its symmetry}

Now we can introduce the BMS $bc$-ghost system to take into account of the Faddeev-Popov determinant
\begin{equation}
  \det (\Delta^{\beta}_{\alpha})  \sim  \int D b D c \exp \left\{ - \int d^2
  \sigma ( c^{\alpha} V^{\gamma} \Delta^{\beta}_{\alpha} b_{\beta \gamma})
  \right\}.
\end{equation}
Here we want the action for the $bc$-ghost to have the symmetry of diffeomorphism, which requires that the whole weight must be $-1$. We choose to include an additional vector density $V$ in the action, since the operator $\Delta^\beta_\alpha$ already has a weight $-\frac{1}{2}$. Then $c^{\alpha}$ is a fermionic vector and $b_{\beta \gamma}$ is a fermionic  tensor, and they form  a multiplet, as we will show in section \ref{multiplet}.

Another problem is that we only need two independent components for $b$-ghost. A simple way to solve the problem is making the tensor $b$ symmetric, i.e. $b_{\alpha\beta}=b_{\beta\alpha}$. As  $b_{11}$ component does not appear in the action, there are exactly two indenpendent components in $b$ ghost. Note that  we cannot make the tensor $b$ traceless because the traceless condition cannot be preserved under the BMS transformations. For convenience, we define
\begin{equation}
  b^0 \equiv b_{00}, \qquad b^1 \equiv b_{10} = b_{01}.
\end{equation}

With the $bc$-ghost,  the partition function becomes
\begin{equation}
  Z = \int D b D c D X e^{i S (V, X, b, c)},
\end{equation}
where
\begin{equation}
\begin{aligned}
  S (V, X, b, c) & \equiv \frac{1}{2 \pi} \int d^2 \sigma \left[ V^{\alpha}
  V^{\beta} \partial_{\alpha} X \cdot \partial_{\beta} X + 2 i c^{\alpha}
  V^{\gamma} \left( \delta^{\beta}_{\quad \alpha} V^{\rho} \partial_{\rho} -
  \partial_{\alpha} V^{\beta} - \frac{1}{2} V^{\beta} \partial_{\alpha}
  \right) b_{\beta \gamma} \right]\\
  & = \frac{1}{2 \pi} \int d^2 \sigma [\partial_0 X \cdot \partial_0 X + i
  (c^0 \partial_0 b^0 - c^1 \partial_1 b^0 + 2 c^1 \partial_0 b^1)].
\end{aligned}\label{inhomogenousbcaction}
\end{equation}
It is invariant under the BMS transformation
\begin{equation}
\begin{aligned}
  \delta_\xi X & = \xi^{\beta} \partial_{\beta} X,\\
  \delta_\xi c^{\alpha} & = - c^{\beta} \partial_{\beta} \xi^{\alpha} +
  \xi^{\beta} \partial_{\beta} c^{\alpha}, \\
  \delta_\xi b_{\alpha \beta} & = b_{\alpha \rho} \partial_{\beta} \xi^{\rho} +
  b_{\rho \beta} \partial_{\alpha} \xi^{\rho} + \xi^{\gamma} \partial_{\gamma}
  b_{\alpha \beta},
\end{aligned}
\end{equation}
or more explicitly,
\begin{equation} \label{trans law}
\begin{aligned}
  \delta_\xi X & =  \xi^{\beta} \partial_{\beta} X,\\
  \delta_\xi c^0 & =  - c^0 \partial_0 \xi^0 - c^1 \partial_1 \xi^0 + \xi^{\beta}
  \partial_{\beta} c^0,\\
  \delta_\xi c^1 & =  - c^1 \partial_1 \xi^1 + \xi^{\beta} \partial_{\beta} c^1,\\
  \delta_\xi b^0 & =  2 b^0 \partial_0 \xi^0 + \xi^{\beta} \partial_{\beta} b^0,\\
  \delta_\xi b^1 & =  b^1 \partial_1 \xi^1 + b^0 \partial_1 \xi^0 + b^1
  \partial_0 \xi^0 + \xi^{\beta} \partial_{\beta} b^1.
\end{aligned}
\end{equation}
The corresponding Noether current and the conserved charge are, respectively,
\begin{equation}
\begin{aligned}
  j^{\mu} & =  \frac{\partial L}{\partial (\partial_{\mu} \Phi)} \delta_\xi
  \Phi - \xi^{\mu} L,\\
  Q  =  \int d \sigma j^0 &= \frac{1}{2\pi} \int d \sigma [\partial_0 X \cdot \partial_0 X - 2 i \partial_1 (c^1
  b_{00}) + i c^1 \partial_1 b_{00}] \xi^0 \\
  &\quad + [2 \partial_0 X \cdot \partial_1 X - 2 i \partial_1 (c^0 b_{00}) + i
  c^0 \partial_1 b_{00} - 4 i \partial_1 (c^1 b_{01}) + 2 i c^1 \partial_1
  b_{01}] \xi^1.
\end{aligned}
\end{equation}
From $Q$ one can read the stress-tensor,  
\begin{equation}
  Q =\int d \sigma (T_1 \xi^1 + T_2 \xi^0)
\end{equation}
with
\begin{equation}
\begin{aligned}
 2\pi T_1 & = 2\partial_0 X \cdot \partial_1 X - i[2 \partial_1 (c^0)
  b_{00} + c^0 \partial_1 b_{00} + 4 \partial_1 (c^1) b_{01} + 2 c^1
  \partial_1 b_{01}],\\
 2\pi T_2 & = \partial_0 X \cdot \partial_0 X - i  [2
  \partial_1 (c^1) b_{00} + c^1 \partial_1 b_{00}].
\end{aligned}
\end{equation}
In addition, one can check that any field $\Phi$ fulfills the equation
\begin{equation}
[Q,\Phi]=i \delta_{\xi} \Phi.
\end{equation}

\subsubsection{BMS algebra}
From the action \eqref{inhomogenousbcaction}, we get the equations of motion
\begin{equation}
\begin{aligned}
  0=\ddot{X}^{\mu},&\hspace{2ex}
  0=\partial_0 c^0 - \partial_1 c^1,&\hspace{2ex}0=\partial_0 c^1,\\
  0=\partial_0 b^0,&\hspace{2ex}
  0=\frac{1}{2} \partial_1 b^0 - \partial_0 b^1,&
\end{aligned}
\end{equation}
which allow us to read the mode expansions 
\begin{equation}
\begin{aligned}
  X^\mu & =  x^\mu + \frac{1}{2} p^\mu \tau + \frac{i}{2} \sum_{n \neq 0} \frac{1}{n} (A^\mu_n - i n
  \tau B^\mu_n) e^{- i n \sigma},\\
  c^1 & =  \sum_n c_n e^{- i n \sigma},\\
  c^0 & =  \sum_n (\tilde{c}_n - i n \tau c_n) e^{- i n \sigma},\\
  b^0 & =  \sum_n \tilde{b}_n e^{- i n \sigma},\\
  b^1 & =  \frac{1}{2} \sum_n (b_n - i n \tau \tilde{b}_n) e^{- i n \sigma}.\\
\end{aligned}
\end{equation}
From canonical quantization, we get the (anti-)commutation relations as
\begin{equation}
\begin{aligned}
  \left[x^\mu, p^\nu\right]  =  i\delta^{\mu\nu}, &\hspace{3ex}
  [A^{\mu}_m, B^{\nu}_n] =  2 m \eta^{\mu \nu} \delta_{m + n},\\
  \{ \tilde{b}_n, \tilde{c}_m \}  =  \delta_{m + n},&\hspace{3ex}
  \{ b_n, c_m \}  =  \delta_{m + n}.
\end{aligned}
\end{equation}
In addition, we can re-combine the coefficients in the mode expansions of $X^\mu$ by
\begin{equation}
\begin{aligned}
&C^\mu_0=\tilde{C}^\mu_0=\frac{1}{2}B^\mu_0=\frac{1}{2}p^\mu,\\
&C^\mu_n=\frac{1}{2}(A^\mu_n+B^\mu_n),\\
&\tilde{C}^\mu_n=\frac{1}{2}(-A^\mu_{-n}+B^\mu_{-n}), 
\end{aligned}
\end{equation}
then we have
\begin{equation}
X^\mu  = x^\mu + C^\mu_0 \tau + \frac{i}{2} \sum_{n \neq 0} \frac{1}{n} [(C^\mu_n -
  \tilde{C}^\mu_{- n}) - i n \tau (C^\mu_n + \tilde{C}^\mu_{- n})] e^{- i n \sigma}, 
  \end{equation}
  with 
 \begin{equation} 
[C^{\mu}_m, C^{\nu}_n] =  m \eta^{\mu \nu} \delta_{m + n}.
\end{equation}

Substituting the mode-expansion into the expression for $T_1$ and $T_2$, we find 
\begin{equation}
\begin{aligned}
 2\pi T_1  = & 2\partial_0 X \cdot \partial_1 X - i [2 \partial_1 (c^0)
  b_{00} + c^0 \partial_1 b_{00} + 4 \partial_1 (c^1) b_{01} + 2 c^1
  \partial_1 b_{01}]\\
   = & \frac{1}{2} \sum_m \sum_{n \neq 0} B_m \cdot (A_n - i n \tau B_n)
  e^{- i (m + n) \sigma}\\
   & - \sum_m \sum_n (2 m + n) [(\tilde{c}_m - i m \tau c_m)
  \tilde{b}_n+c_m (b_n - i n \tau \tilde{b}_n)] e^{- i (m + n) \sigma},\\
 2\pi  T_2 = &  \partial_0 X \cdot \partial_0 X - i  [2
  \partial_1 (c^1) b_{00} + c^1 \partial_1 b_{00}]\\
   = & \frac{1}{4} \sum_m \sum_n B_m \cdot B_n e^{- i (m + n) \sigma} -   \sum_m \sum_n (2 m + n) c_m \tilde{b}_n e^{- i (m + n).
  \sigma}
\end{aligned}
\end{equation}
By using the Fourier transformation, we obtain the generators
\begin{equation}\nonumber
\begin{aligned}
  L_n & = \int d \sigma (T_1 + i n \tau T_2) e^{i n \sigma}\\
  & =  \frac{1}{2} \sum_m A_m \cdot B_{n - m} + \sum_m (n - m) (b_{n + m}
  c_{- m} + \tilde{b}_{n + m} \tilde{c}_{- m}),\\
  M_n & = \int d \sigma T_2 e^{i n \sigma}\\
  & =  \frac{1}{4} \sum_m B_m \cdot B_{n - m} + \sum_m (n - m)
  \tilde{b}_{n + m} c_{- m},
\end{aligned}
\end{equation}
which generate  the BMS algebra
\begin{equation}
\begin{aligned}
  &&[L_m, L_n]  = & (m - n) L_{m + n} + A_L (m) \delta_{m + n,0},\\
  &&[L_m, M_n]  = & (m - n) M_{m + n} + A_M (m) \delta_{m + n,0},\\
  &&[M_m, M_n]  = & 0.
\end{aligned}
\end{equation}
Here $A_L$ and $A_M$ are the anomaly terms, which include the contributions from the central charges and the normal-ordering terms.

If we want all the anomalies to vanish, the critical-dimension and
the normal-ordering constant must satisfy certain conditions. This is only available for the flipped vacuum and the
induced vacuum,  because in the oscillating vacuum, most generators acting on the vacuum give
infinite number of states and there is no appropriate way to regulate the resulting infinity.

In the induced vacuum, which is annihilated by all non-zero $B$ modes, i.e.
\begin{equation}
   B_n | 0 \nobracket \rangle_{\text{Ind}} =  0, \qquad n \neq 0,
\end{equation}
there is no
constraint on the critical dimension, which means
\begin{equation}
A_L (m) = 0, \qquad A_M(m)=0
\end{equation}
if only the normal-ordering term $a = 0$.

For the flipped vacuum which is in the highest-weight representation, it is defined by 
\begin{equation}
\begin{aligned}
  A_n | 0 \nobracket \rangle \nobracket = B_n | 0 \nobracket \rangle
  \nobracket & =  0, \qquad n \geqslant 0,\\
  c_n | 0 \nobracket \rangle \nobracket = b_n | 0 \nobracket \rangle
  \nobracket & =  0, \qquad n > 0,\\
  L_n | 0 \nobracket \rangle \nobracket = M_n | 0 \nobracket \rangle
  \nobracket & =  0, \qquad n \geqslant 0.\\
\end{aligned}
\end{equation}
Then we have the anomaly terms
\begin{equation}
\begin{aligned}
  & A_L (m) = \frac{D}{6} (m^3 - m)- \frac{1}{3} (13 m^3 -
  m),\\
  &  A_M (m) = 0,
\end{aligned}
\end{equation}
where $D$ is the dimension of the target-space of the string.
Introducing  the normal-ordering term $a_L$, we find
\begin{equation}
\begin{aligned}
  L_m & \rightarrow  L_m  - a_L \delta_m,\\
  A_L (m) & \rightarrow  \frac{D}{6} (m^3 - m) - \frac{1}{3} (13 m^3 - m) + 2
  a_L m.
\end{aligned}
\end{equation}
and the vanishing of $A_L (m)$ requires $D = 26$ and $a_L = 2$, which are the same as the ones in the tensile string.

\subsection{BMS \texorpdfstring{$bc$}~-ghost from the UR limit}

In this subsection, we show how to obtain the tensionless $bc$ ghost by taking inhomogeneous ultra-relativistic(UR) limit on the tensile string.
In the usual tensile string theory including the $bc$ ghost, we have the action
\begin{equation}
\begin{aligned}
  S & =  - \frac{1}{2 \pi} \int d^2 \sigma \sqrt{- g} g^{\alpha \beta}
  (\partial_{\alpha} X \partial_{\beta} X + i c^{\gamma} \nabla_{\alpha}
  b_{\beta \gamma})\\
  & = \frac{1}{2 \pi} \int d^2 \sigma [(\partial_t X \cdot \partial_t X -
  \partial_1 X \cdot \partial_1 X) + i (c^{\gamma} \partial_t b_{t \gamma} -
  c^{\gamma} \partial_1 b_{1 \gamma})],
\end{aligned}
\end{equation}
keeping in mind that $b$ is symmetric and traceless,
\begin{equation}
  b_{t t}  =  b_{11},\hspace{3ex}
  b_{t 1}  =  b_{1 t}.
\end{equation}
The $bc$ ghosts in the tensile string theory have the mode expansions as\footnote{Notations: we use the Gothic characters to stand for the mode coefficients of the original tensile string; we use the index label $\tau$ or $0$ for the tensionless string and the label $t$ for the tensile string.}
\begin{equation}
\begin{aligned}
  c^t&=\sum_n \left[\mathfrak{c}_n e^{- i n (t+\sigma)}+\tilde{\mathfrak{c}}_n e^{- i n (t-\sigma)} \right], \\
  c^1&=\sum_n \left[\mathfrak{c}_n e^{- i n (t+\sigma)}-\tilde{\mathfrak{c}}_n e^{- i n (t-\sigma)} \right], \\
  b_{1 t}&=\frac{1}{2} \sum_n \left[\mathfrak{b}_n e^{- i n (t+\sigma)}-\tilde{\mathfrak{b}}_n e^{- i n (t-\sigma)}, \right] \\
  b_{1 1}&=\frac{1}{2} \sum_n \left[\mathfrak{b}_n e^{- i n (t+\sigma)}+\tilde{\mathfrak{b}}_n e^{- i n (t-\sigma)} \right], \\
\end{aligned}
\end{equation}
with the anti-commutation relations
\begin{equation}
\begin{aligned}
\{\mathfrak{c}_m,\mathfrak{b}_n\}=\delta_{m+n}, \\
\{\tilde{\mathfrak{c}}_m,\tilde{\mathfrak{b}}_n\}=\delta_{m+n}.
\end{aligned}
\end{equation}
We may take the following UR limit
\begin{equation}
\begin{aligned}
  \epsilon \rightarrow 0,\hspace{3ex} & \xi^t  =  \epsilon \xi^\tau\equiv \epsilon \xi^0, \\
  t  =  \epsilon \tau,\hspace{3ex} &
  \partial_t  =  \frac{\partial_{\tau}}{\epsilon} \equiv
  \frac{\partial_0}{\epsilon}, 
\end{aligned}
\end{equation}
and find that the action becomes
\begin{equation}
\begin{aligned}
  S_X & =  \frac{1}{2 \pi} \int d^2 \sigma \epsilon^{- 1} (\partial_0 X \cdot
  \partial_0 X - \epsilon^2 \partial_1 X \cdot \partial_1 X),\\
  S_g & =  \frac{i}{2 \pi} \int d^2 \sigma \epsilon (c^{\gamma} \partial_t
  b_{t \gamma} - c^{\gamma} \partial_1 b_{1 \gamma})\\
  & =  \frac{i}{2 \pi} \int d^2 \sigma (c^t \partial_0 b_{11} + c^1
  \partial_0 b_{t 1} - \epsilon c^t \partial_1 b_{t 1} - \epsilon c^1
  \partial_1 b_{11}).
\end{aligned}
\end{equation}


Before considering the UR limit of $bc$ ghosts, we should consider the BMS transformations in the limit carefully. The 
original diffeomorphism takes the form
\begin{equation}
  \partial_+ \xi^-  = \partial_- \xi^+ = 0
\end{equation}
where
\begin{equation}
\begin{aligned}
  \partial_+  =  \frac{\partial_t + \partial_1}{\sqrt{2}},\hspace{3ex}&
  \partial_-  =  \frac{\partial_t - \partial_1}{\sqrt{2}},\\
  \xi^+ = \frac{\xi^t + \xi^1}{\sqrt{2}},\hspace{3ex}&
  \xi^-  = \frac{\xi^t - \xi^1}{\sqrt{2}}.
\end{aligned}
\end{equation}
Under the UR limit, the transformation conditions  become
\begin{equation}
\begin{aligned}
  \partial_0 \xi^0 & =  \partial_1 \xi^1,\\
  \partial_0 \xi^1 & =  \epsilon^2 \partial_1 \xi^0 = O (\epsilon^2).
\end{aligned}
\end{equation}
Similarly, considering the BMS algebra as the UR limit of two copies of Virasoro algebra carefully, we find
\begin{equation}
    [M_m,M_n]=\epsilon^2 [L_m,L_n]=O(\epsilon^2).
\end{equation}
This suggests that we should not neglect $O (\epsilon^2)$ terms in taking the UR limit.

\subsubsection*{Inhomogenious limit}

There are different options for the $bc$-ghost fields in the UR limit. Let us first consider the inhomogenious limit, which requires that 
\begin{equation}
\begin{aligned}
  c^t & =  \epsilon c^0,\\
  b_{t 1} = b_{1 t} & =  \epsilon b_{01} = \epsilon b_{10},
\end{aligned}
\end{equation}
with other components remaining the same. 
If we define
\begin{equation}
  b^0  \equiv  b_{11},\hspace{3ex}
  b^1  \equiv  \frac{1}{2} b_{01}, 
\end{equation}
we reproduce the theory discussed before through intrinsic analysis
\begin{equation}
  S  \rightarrow  \frac{1}{2 \pi} \int d^2 \sigma [\partial_0 X \cdot
  \partial_0 X + i (c^0 \partial_0 b^0 - c^1 \partial_1 b^0 + 2 c^1 \partial_0
  b^1)]. 
\end{equation}
The transformation rules for the fields can be obtained by taking the UR limit as well. 
Comparing the mode expansions before and after taking the UR limit, we find
\begin{equation}\label{rela-imh}
\begin{aligned}
  \mathfrak{c}_n = c_n+\epsilon\tilde{c}_{n},\hspace{3ex}&
  \mathfrak{b}_n = (b_n+\tilde{b}_{n}/\epsilon)/2,\\
  \tilde{\mathfrak{c}}_n = -c_{-n}+\epsilon \tilde{c}_{-n},\hspace{3ex}&
  \tilde{\mathfrak{b}}_n = (-b_{-n}+\tilde{b}_{-n}/\epsilon)/2.\\
\end{aligned}
\end{equation}

\subsubsection*{Homogeneous limit}

Next, we turn to another option for the $bc$-ghost fields in the UR limit. In this so-called homogeneous limit, all the fields rescale in the same way
\begin{equation}
  c^t = c^0,\hspace{3ex}
  b_{t 1} = b_{1 t}  =  b_{01} = b_{10}, 
\end{equation}
and the action for the ghost is 
\begin{equation}
  S_g  =  \frac{i}{2 \pi} \int d^2 \sigma \epsilon (c^0 \partial_0 b_{11} + c^1
  \partial_0 b_{01}).
\end{equation}
With the definition
\begin{equation}
  b^0  \equiv  b_{11},\hspace{3ex}
  b^1  \equiv  b_{01},
\end{equation}
we obtain a theory with action
\begin{equation}
  S =  \frac{1}{2 \pi} \int d^2 \sigma [\partial_0 X \cdot
  \partial_0 X + i (c^0 \partial_0 b^0 + c^1 \partial_0 b^1)]
\end{equation}
which is quite different from the theory we studied before. 
From the equations of motion, we can get the mode expansions of the fields. For the field $X^\mu$, it is the same as the inhomogeneous case. But the mode expansion for the ghost fields are different, 
\begin{equation}
\begin{aligned}
  c^0  =  \sum_n c_n e^{- i n \sigma},\hspace{3ex}&
  c^1  =  \sum_n \tilde{c}_n e^{- i n \sigma},\\
  b^0  =  \sum_n \tilde{b}_n e^{- i n \sigma},\hspace{3ex}&
  b^1  =  \sum_n b_n e^{- i n \sigma}.
\end{aligned}
\end{equation}
The canonical quantization leads to the anti-commutation relations
\begin{equation}
  \{ b_n, c_m \}  =  \delta_{m + n},\hspace{3ex}
  \{ \tilde{b}_n, \tilde{c}_m \}  =  \delta_{m + n}. 
\end{equation}
Comparing the mode expansions before and after the UR limit, we find
\begin{equation}\label{rela-h}
\begin{aligned}
  \mathfrak{c}_n = c_n+\tilde{c}_{n},\hspace{3ex}& \mathfrak{b}_n = (b_n+\tilde{b}_{n})/2,\\
  \tilde{\mathfrak{c}}_n = c_{-n}-\tilde{c}_{-n},\hspace{3ex}&
  \tilde{\mathfrak{b}}_n = (b_{-n}-\tilde{b}_{-n})/2.
\end{aligned}
\end{equation}
With the mode expansions, we obtain the BMS generators
\begin{equation}
\begin{aligned}
  L_n &  =  \frac{1}{2} \sum_m A_m \cdot B_{n - m} + \sum_m (n - m) (b_{n + m}
  c_{- m} + \tilde{b}_{n + m} \tilde{c}_{- m}),\\
  M_n  & =  \frac{1}{4} \sum_m B_m \cdot B_{n - m}.
\end{aligned}
\end{equation}
The Virasoro generators take the same forms as the ones in the inhomogeneous case, while the generators $M_n$ are slightly different. We find the same critical dimension and the normal-ordering constant. 

\subsubsection*{BRST charge}

Here we demonstrate that only the inhomogenous $bc$ ghost is the correct ghost system for the tensionless string. Actually, the intrinsic analysis implies that there must be a derivative with respect to the coordinate $\sigma$ in the action for the ghosts, but the homogeneous $bc$-ghost fields do not contain such a term. More explicitly, we can check the BRST charge, and we find  that only in the inhomogenous case, the UR limit of the BRST charge of original tensile string theory is meaningful.

Let us start from the  BRST charge of original tensile string
\begin{equation}
\begin{aligned}
Q_B=&\sum_m(\mathcal{L}_{m}^{(X)}-a \delta_m)\mathfrak{c}_{-m}-\frac{1}{2}\sum_{m,n}(m-n):\mathfrak{c}_{-m}\mathfrak{c}_{-n}\mathfrak{b}_{m+n}:+c.c.\\
=&\sum_m :(\mathcal{L}_{m}^{(X)}+\frac{1}{2}\mathcal{L}_{m}^{(ghost)}-a \delta_m)\mathfrak{c}_{-m}:+c.c., 
\end{aligned}
\end{equation}
which satisfies the relation
\begin{equation}\label{tedi}
    \{Q_B,Q_B\}=\sum_{m,n} \left([\mathcal{L}_m,\mathcal{L}_n]-(m-n)\mathcal{L}_{m+n}\right)\mathfrak{c}_{-m}\mathfrak{c}_{-n}+c.c..
\end{equation}
Then we can consider different limits. It turns out that the Virasoro generators are related to the generators of BMS algebra in the following way,
\begin{equation}\label{urrel}
\begin{aligned}
\mathcal{L}_n^{(\text{fields})}&=\frac{L_n^{(\text{fields})}+M_n^{(\text{fields})}/\epsilon}{2},\\
\bar{\mathcal{L}}_n^{(\text{fields})}&=\frac{-L_{-n}^{(\text{fields})}+M_{-n}^{(\text{fields})}/\epsilon}{2}.\\
\end{aligned}
\end{equation}

In the homogenous case, the relation (\ref{urrel}) holds only for the fields  $X^\mu$. But in the inhomogenous case, the relation (\ref{urrel}) holds not only for the fields $X^\mu$ but also for the $bc$-ghost generators. After calculations, we obtain a finite BRST charge for the inhomogeneous case
\begin{equation}
Q_B=\sum_m :\left(L_m^{(X)}+\frac{1}{2}L_m^{(ghost)}-a_L \delta_m\right) c_{-m}:+\sum_m :\left(M_m^{(X)}+\frac{1}{2}M_m^{(ghost)}-a_M \delta_m\right)\tilde{c}_{-m}:,
\end{equation}
where $a_L$ and $a_M$ are introduced in the same way as in (\ref{urrel}). 
And the ghost number operator becomes
\begin{equation}
U=\sum_m :(c_{-m}b_m+\tilde{c}_{-m}\tilde{b}_m):
\end{equation}
This BRST charge can also be expressed in terms of BMS fields defined on the null plane as
\begin{equation}
Q_B=\frac{1}{2\pi i}\oint d x :\left[\left(T^{(X)}+\frac{1}{2}T^{(ghost)}-\left(a_L-2 a_M \frac{y}{x}\right) x^{-2}\right)c^1+\left(M^{(X)}+\frac{1}{2}M^{(ghost)}-a_M x^{-2}\right)c^0\right]:
\end{equation}
Then we can use the operator product expansion(OPE) to check its nilpotency which is obviously correct if the anomaly vanishes since the UR limit works well. In fact, the UR limit of eq.(\ref{tedi}) gives us
\begin{equation}
\begin{aligned}
        \{Q_B,Q_B\}=&\sum_{m,n} \left([L_m,L_n]-(m-n)L_{m+n}\right)c_{-m}c_{-n}\\
        &+\sum_{m,n} \frac{1}{2}\left([L_m,M_n]+[M_m,L_n]-2(m-n)M_{m+n}\right)(\tilde{c}_{-m}c_{-n}+c_{-m}\tilde{c}_{-n}).
\end{aligned}
\end{equation}

In the homogenous case, the relation (\ref{urrel}) does not hold for the $bc$-ghost generators. In this case, after some calculations, we obtain a divergent BRST charge
\begin{equation}
Q_B=\sum_m M_m^{(X)}c_{-m}/\epsilon+O(\epsilon^0),
\end{equation}
which suggest that the homogenenous $bc$-ghost does not make sense. 

\section{Tensionless superstrings and BMS \texorpdfstring{$\beta\gamma$}~ ghosts}\label{section3}

    In the previous section, we have discussed bosonic tensionless string and the corresponding $bc$ ghosts, while in this section we turn to the  fermionic tensionless strings and the corresponding $\beta\gamma$ ghosts. We discuss in detail in section \ref{subsec:ActionofFermionicString} the taking-limit procedure from tensile fermionic string, and find that different rescalings of the fermion field components result in homogeneous and inhomogeneous fermionic string actions. For every type of fermionic string, we have two different rescalings of gravitino field components, and we obtain four tensionless superstring actions. Using the Faddeev-Popov trick, we introduce the corresponding $\beta\gamma$-ghost field actions to fix the gauge degrees of freedom of the gravitino field. Further in section \ref{subsec:SuperstringCriticalDim}, we use the canonical quantization to read the critical dimensions of the four types  of tensionless superstrings. Strangely, the critical dimensions of two types of tensionless superstring are not integers.

\subsection{Tensionless fermionic strings and \texorpdfstring{$\beta\gamma$}~-ghost}\label{subsec:ActionofFermionicString}
    In the literature, the tensionless fermionic string was constructed by taking the UR limit from relativistic superstring \cite{Bagchi:2016yyf,Bagchi:2017cte,Bagchi:2018wsn}. As in the bosonic case, we can get the $\beta\gamma$ ghost of the tensionless fermionic string by taking the tensionless limit from tensile $\beta\gamma$ ghost. However, similar to the $bc$-ghost case discussed before, there is ambiguity in determining the scaling behavior of the ghost fields. Thus we prefer to construct the $\beta\gamma$ ghost intrinsically, where we can uniquely determine the ghost Lagrangian. \par
    
    In this section, we try to construct $\beta\gamma$-ghost fields intrinsically by cancelling the gravitino degrees of freedom. It has been found that in taking the tensionless limit, two different rescalings of the fermions lead to two kinds of fermions, i.e. homogeneous and inhomogeneous ones, and two rescalings of the gravitinos. Together, there are four kinds of $\beta\gamma$-ghost Lagrangians.
    
     \par
    
    Let us first briefly review the construction of $\beta\gamma$ ghost for the tensile fermionic  string. The Lagrangian of the superstring in a curved spacetime is 
    \begin{equation}
        \mathcal{L}=\eta^{ab}e^\alpha_a e^\beta_b \partial_\alpha X\partial_\beta X + 2 \bar{\chi}_\beta \uline{\rho}^\alpha \uline{\rho}^\beta \psi \partial_\alpha X + \frac{1}{2}\bar{\psi}\psi\bar{\chi}_\alpha\uline{\rho}^\beta\uline{\rho}^\alpha\chi_\beta -i \bar{\psi}\uline{\rho}^\alpha\nabla_\alpha\psi,
    \end{equation}
    where the Greek letters $\alpha, \beta, ... =0, 1$ are the worldsheet indices. The gravitino field $\chi_\alpha$ actually carries the spinor index $\chi_\alpha=\chi_{\alpha A}$ with the capital letters $A,B, ... =0, 1$ being the spinor indices, and thus it contains four components, i.e. $\chi_{00}$, $\chi_{01}$, $\chi_{10}$, and $\chi_{11}$. There are two fermionic symmetries for this Lagrangian, one being the super symmetry(SUSY) transformation 
    \begin{equation}\label{eq:SUSYTransformation}
        \begin{aligned}
            \delta_{\text{SUSY}} X &= \bar{\epsilon}\psi, \qquad 
            &\delta_{\text{SUSY}}\psi &= -i\uline{\rho}^\alpha\epsilon (\partial_\alpha X - \bar{\psi}\chi_\alpha),\\
            \delta_{\text{SUSY}} e^\alpha_a &= -\eta_{ab}\bar{\chi}_\gamma\uline{\rho}^\alpha\uline{\rho}^\gamma\rho^b \epsilon, \qquad 
            &\delta_{\text{SUSY}} \chi_{\alpha} &= \nabla_\alpha \epsilon ,\\
        \end{aligned}
    \end{equation}
    and the other being super-conformal transformation
    \begin{equation}
        \delta_{\text{SC}} X = 0, \qquad 
        \delta_{\text{SC}}\psi = 0, \qquad 
        \delta_{\text{SC}} e^\alpha_a =0, \qquad 
        \delta_{\text{SC}} \chi_{\alpha} = i h_{\alpha\beta}\uline{\rho}^\beta\eta,
    \end{equation}
    where $\eta$ is a Majorana fermion. The total number of degrees of freedom in these transformations is four, which just fixes the degrees of freedom of $\chi$. The fixing is done by inserting in the path integral the identity
    \begin{equation}
        1=\int \mathcal{D}\epsilon \mathcal{D}\eta ~ \frac{\delta \chi}{\delta(\epsilon,\eta)} ~ \delta(\chi-\chi^\prime),
    \end{equation}
    where $\chi^\prime$ is the transformed gravitino under a SUSY transformation generated by $\epsilon$ and a super-conformal transformation generated by $\eta$, and similar for $X^\prime$ and $\psi^\prime$ in the following equations. The Jacobian $\frac{\delta \chi}{\delta(\epsilon,\eta)}$ can be rewritten in terms of the path-integral of the ghost field(s) $f_\text{ghost}$:
    \begin{equation}
        \frac{\delta \chi}{\delta(\epsilon,\eta)} = \int [\mathcal{D} f_\text{ghost}] \exp{iS_\text{ghost}}.
    \end{equation}
    Thus the generating function of the superstring is
    \begin{equation}\label{eq:FadeevPopovMeasure}
        \begin{aligned}
            Z&=\int \mathcal{D}X \mathcal{D}\psi \mathcal{D}\chi \exp{ i S_\text{matter}[X,\psi,\chi]}\\
                &=\int \mathcal{D}X \mathcal{D}\psi \mathcal{D}\chi \mathcal{D}\epsilon \mathcal{D}\eta [\mathcal{D} f_\text{ghost}] ~ \delta(\chi-\chi^\prime) \exp{ i S_\text{matter}[X,\psi,\chi] + iS_\text{ghost}} \\
                &=\int \mathcal{D}\epsilon \mathcal{D}\eta \int \mathcal{D}X^\prime \mathcal{D}\psi^\prime[\mathcal{D} f_\text{ghost}] ~ \exp{ i S_\text{matter}[X^\prime,\psi^\prime,\chi^\prime] + iS_\text{ghost}} \\
                &\sim \int \mathcal{D}X \mathcal{D}\psi[\mathcal{D} f_\text{ghost}] ~ \exp{ i S_\text{matter}[X,\psi,\chi^\prime] + iS_\text{ghost}} ,\\
        \end{aligned}
    \end{equation}
    where we drop an infinite overall coefficient. The Faddeev-Popov determinant  arose in fixing $\chi$ is compensated by introducing the $\beta\gamma$ fields into the action. For tensionless fermionic string in different limits, the discussion will be slightly different, as we will show case by case.

    In \cite{Lindstrom:1993yb},  the tensionless limit of string theory was proposed. For the fermion with two components, there is a degree of freedom in taking the tensionless limit, as the relativistic rescalings on two components could be same or different. Different rescalings correspond to homogeneous and inhomogeneous sectors, respectively. Here we apply the ILST trick for the fermionic string. The action of a tensile string is
    \begin{equation}\label{eq:TensileStringAction}
        S=-\frac{T}{2}\int d^2 \sigma \sqrt{h} ~ \mathcal{L},
    \end{equation}
    where $h^{\alpha\beta}$ is the inverse of the metric on the worldsheet, and $h=\abs{\det h_{\alpha\beta}}$. Obviously,  taking $T\to 0$ limit directly in \eqref{eq:TensileStringAction} gives nothing interesting. The ILST trick \cite{Lindstrom:1993yb} is to introduce two Lagrange multipliers $\lambda$ and $\rho$ such that the metric becomes
    \begin{equation}
        h^{\alpha\beta}=\begin{pmatrix}
            -1 & \rho\\
            \rho & 4\lambda^2 T^2 -\rho^2
        \end{pmatrix}, \qquad \sqrt{h}=\frac{1}{2\lambda T}=\tilde{T}^{-1},
    \end{equation}
    where we denote $\tilde{T}=2\lambda T$ for simplicity. Thus the action becomes
    \begin{equation}
        S=-\frac{1}{4\lambda}\int d^2 \sigma \mathcal{L},
    \end{equation}
    and we are allowed to take the $T\to 0$ limit. 
    
    To consider the fermions, we further introduce the inverse of the zweibein $e^\alpha_a$, which have one Lorentzian degree of freedom parameterized by $\abs{\cosh{\theta}}=\abs{a}\ge 1$,
    \begin{equation}
        \begin{aligned}
            e^\alpha_0 &=(a,  &-a\rho+\sqrt{a^2-1}~\tilde{T}),\\
            e^\alpha_1 &=(\sqrt{a^2-1},  &-\sqrt{a^2-1}~\rho+a\tilde{T}),\\
        \end{aligned}
        \qquad\qquad \eta^{ab} e^\alpha_a e^\beta_b=h^{\alpha\beta}.
    \end{equation}
   Note that we only define the inverse of the zweibein, not the zweibein itself, since when we take the tensionless limit $\tilde{T}\to0$, the inverse of the zweibein is degenerate. The Dirac gamma matrices $\uline{\rho}^\alpha$ in curved space are related to the ones $\rho^a$ in flat spacetime by
    \begin{equation}
        \uline{\rho}^\alpha=e^\alpha_a \rho^a, 
        \qquad \text{with}\quad \rho^0=\begin{pmatrix} 0 & -i\\ i & 0\end{pmatrix}, 
        \qquad \rho^1=\begin{pmatrix} i & 0\\ 0 & -i\end{pmatrix}.
    \end{equation}
    The tensile fermionic Lagrangian is 
    \begin{equation}\label{eq:LagrangianofFermion}
        \mathcal{L}_f=-i\bar{\psi}^\mu\uline{\rho}^\alpha\partial_\alpha\psi_\mu, \qquad  \quad \bar{\psi}^\mu=(\psi^\mu)^\dagger\rho^0,
    \end{equation}
   where the fermion is Majorana fermion, i.e. $\psi_A^\dagger=\psi_A$ with capital Latin letters $A,B,...$ label the spinor index. In the following, we omit the spacetime index $\mu$ and set the Lagrangian multiplier $\rho=0$ for simplicity, and thus have
    \begin{equation}
        \uline{\rho}^0=\begin{pmatrix} i \sqrt{a^2-1} & -i a\\ i a & -i \sqrt{a^2-1}\end{pmatrix}, 
        \qquad \uline{\rho}^1=\tilde{T}\begin{pmatrix} i a & -i \sqrt{a^2-1}\\ i \sqrt{a^2-1} & -i a\end{pmatrix}.
    \end{equation}
    
    
    In taking the tensionless limit $\tilde{T}\to0$, we should keep tracking of the rescaling of the components of the fermion,
    \begin{equation}
        \psi_0\to\tilde{T}^{s_0}\psi_0, \quad \psi_1\to\tilde{T}^{s_1}\psi_1.
    \end{equation}
    It should be noticed that there is also a rescaling factor for the Lorentzian rotation generator in spinor representation:
    \begin{equation}
        S_\psi=\frac{\tilde{T}}{2}\begin{pmatrix} 0 & 1\\ 1 & 0\end{pmatrix},
    \end{equation}
    where $S_\psi$ is the representation matrix of the rotation $S$ acting on $\psi$. In order to keep both components of the fermion $\psi$ in the tensionless limit, the scale factors must obey the relation
    \begin{equation}
        |s_0-s_1|\le 1,
    \end{equation}
    which have two independent solutions, corresponding to the homogeneous limit with $|s_0-s_1|=0$ and inhomogeneous limit with $|s_0-s_1|=1$. On the other hand, if we impose $|s_0-s_1|> 1$, one of the fermion component (say $\psi_1$) vanishes in taking the $\tilde{T}\to 0$ limit, and we get a trivial Lagrangian 
    \begin{equation}\label{eq:TrivialLagrangian}
        \mathcal{L}_f^{\text{tri}}= -i \psi_0\partial_0\psi_0.
    \end{equation} 
    This trivial Lagrangian  can also appear in the reduction of homogeneous superstring, and the corresponding $\beta\gamma$ ghost will be discussed later. 
    
\subsubsection*{Homogeneous limit}
    The homogeneous limit corresponds to $s_0=s_1$. We may set $s_0=s_1=0$ to keep the Lagrangian finite in $\tilde{T}\to 0$. Thus we find that the two components of homogeneous fermion are invariant under rotation
    \begin{equation}\label{eq:RepsofHhomoFermion}
        \begin{aligned}
            \left[S,\psi_A\right]&=0.
        \end{aligned}
    \end{equation}
    With the parameter $a=1$, the gamma matrices are
    \begin{equation}\label{eq:HomoGammaMatrices}
        \uline{\rho}^0=\begin{pmatrix} 0 & -i\\ i & 0\end{pmatrix}, 
        \qquad \uline{\rho}^1=\begin{pmatrix} 0 & 0\\ 0 & 0\end{pmatrix},
    \end{equation}
    and the Lagrangian is
    \begin{equation}\label{eq:HomoLagrangian}
        \mathcal{L}_f^{\text{(h)}}= -i \psi_0\partial_0\psi_0 -i \psi_1\partial_0\psi_1.
    \end{equation}
    As the two components of the fermion are independent with each other, it is possible to turn off either one of them, leading to the trivial Lagrangian \eqref{eq:TrivialLagrangian}.  \par
    
    The two components of the fermion in the homogeneous limit are in a singlet representation (\ref{eq:RepsofHhomoFermion}), thus we should set two components of $\epsilon$ in the SUSY transformation also in a singlet representation. Plugging in the gamma matrices (\ref{eq:HomoGammaMatrices}), we know that the $\alpha=1$ components of $\chi$ do not contribute to the Lagrangian and the SUSY transformation. Besides, for the tensionless Lagrangian, the super-conformal transformation is now
    \begin{equation}
        \delta_{\text{SC}} X = 0, \qquad 
        \delta_{\text{SC}}\psi = 0, \qquad 
        \delta_{\text{SC}} e^\alpha_a =0, \qquad 
        \delta_{\text{SC}} \chi_{\alpha=0} = 0.
    \end{equation}
    Thus there are exactly two degrees of freedom in $\chi$ in the Lagrangian to cancel, while there are two degrees of freedom of $\epsilon$ in the SUSY transformation. Using the fact that
    \begin{equation}
        \det \left(\frac{\delta \chi_{\alpha=0}}{\delta \epsilon}\right) \sim \int \mathcal{D}\beta_{00} \mathcal{D}\beta_{01} \mathcal{D}\gamma_0 \mathcal{D}\gamma_1 \exp \left\{ -\frac{1}{2\pi} \int d^2 \sigma \left( \gamma_1\partial_0\beta_{00} - \gamma_0\partial_0\beta_{01} \right)\right\},
    \end{equation}
    we immediately read the  Lagrangian of $\beta\gamma$ ghosts 
    \begin{equation}
        \begin{aligned}
            \mathcal{L}^{\text{(h)}}_{\beta\gamma} &=i h^{\alpha\beta}\bar{\gamma}\nabla_\alpha \beta_{\beta}= \gamma_1\partial_0\beta_{00} - \gamma_0\partial_0\beta_{01},\\
        \end{aligned}
    \end{equation}
    where the second equality is the expression in the flat space gauge. $\gamma_A$ is a bosonic spinor field, and $\beta_{\alpha A}$ is a bosonic vector-spinor field whose $\alpha=1$ components do not contribute. With the bosonic sector together, we have the Lagrangian of homogeneous tensionless superstring:
    \begin{equation}\label{eq:HomogeneousTensionlessSuperStringLagrangian}
        \begin{aligned}
            \mathcal{L}^\text{(h)}=&-\partial_0 X\partial_0 X - i c^0\partial_0 b_{00} +i c^1\partial_1b_{00} -2ic^1\partial_0 b_{01} \\
            &\qquad - i\psi_0\partial_0\psi_0 - i\psi_1\partial_0\psi_1 + \gamma_0\partial_0 \beta_{0} + \gamma_1\partial_0 \beta_{1}.
        \end{aligned}
    \end{equation}

    As mentioned before, we can turn off $\psi_1$ to get a trivial Lagrangian (\ref{eq:TrivialLagrangian}). Correspondingly, we can turn off $\beta_{01}$ and $\gamma_0$ to get its ghost Lagrangian
    \begin{equation}
        \mathcal{L}^{\text{(tri)}}_{\beta\gamma} = \gamma_1\partial_0\beta_{00}.
    \end{equation}
    Thus the full trivial tensionless superstring Lagrangian is
    \begin{equation}\label{eq:TrivialTensionlessSuperStringLagrangian}
        \begin{aligned}
            \mathcal{L}^\text{(tri)}=&-\partial_0 X\partial_0 X - i c^0\partial_0 b_{00} +i c^1\partial_1b_{00} -2ic^1\partial_0 b_{01} - i\psi_0\partial_0\psi_0 + \gamma_0\partial_0 \beta_{0}.
        \end{aligned}
    \end{equation}
    
\subsubsection*{Inhomogeneous limit}
    The inhomogeneous limit corresponds to $|s_0-s_1|= 1$ and we take $s_0=-\frac{1}{2}, s_1=\frac{1}{2}$ to keep the Lagrangian finite in taking $\tilde{T}\to 0$. In this limit, the fermion is in doublet representation under the rotation $S$ 
    \begin{equation}\label{eq:RepsofInhomoFermion}
        \begin{aligned}
            \left[S,\psi_A\right] & =\frac{1}{2}\delta_{1,A}\psi_0.
        \end{aligned}
    \end{equation}
    where $\delta_{1,A}$ is Kronecker delta function. However, the standard treatment of the inverse of the zweibien leads to a trivial Lagrangian. To get the inhomogeneous Lagrangian, we have to do analytic continuation on the parameter $a$ such that it could take any real value $a\in\mathbb{R}$ rather than $|a|\geq 1$. And finally we  take $a=0$ to get 
    \begin{equation}\label{eq:InhomoGammaMatrices}
        \uline{\rho}^0=\begin{pmatrix} -1 & 0\\ 0 & 1\end{pmatrix}, 
        \qquad \uline{\rho}^1=\begin{pmatrix} 0 & 0\\ -1 & 0\end{pmatrix}.
    \end{equation}
    In fact, extending to $a=0$ lead to purely imaginary zweibien and a real gamma matrix $\uline{\rho}^1$. Notice that the flat gamma matrix $\rho^0$ is untouched in this extension and remains still pure imaginary. The extra imaginary factor $i$ introduced by this extension should be taken out by hand, such that for Majorana fermion $\psi$, the Lagrangian is real
    \begin{equation}\label{eq:InhomoLagrangian}
        \mathcal{L}_f^{\text{(ih)}}= -i \psi_1\partial_0\psi_0 -i \psi_0\partial_0\psi_1 + i \psi_0\partial_1\psi_0 .
    \end{equation}
    \par

    Similarly, there is ambiguity in the relative rescalings on different components in $\chi$. There are two different rescalings, corresponding to two representations  of the rotation $S$, 
    \begin{align}
        &\text{vector-doublet: }\chi^{(d)}_{\alpha A} 
            \qquad \left[S,\chi^{(d)}_{\alpha A}\right]=\delta_{1,\alpha}\chi^{(d)}_{0A} + \frac{1}{2}\delta_{1,A}\chi^{(d)}_{\alpha 0},\\
        &\text{vector-singlet: }~\chi^{(s)}_{\alpha A} 
            \qquad \left[S,\chi^{(s)}_{\alpha A}\right]=\delta_{1,\alpha}\chi^{(s)}_{0A}.
    \end{align}
    The corresponding SUSY transformations and super-conformal transformations are different. Let us discuss them case by case. \par

    \textbf{Vector-doublet gravitino} 
    In this case, $\chi^{(d)}_{11}$ dose not contribute to the Lagrangian thus the Lagrangian only contains three $\chi$ components. The Lagrangian in the flat gauge is 
    \begin{equation}
        \begin{aligned}
            \mathcal{L}^\text{(d-ih)}_\text{matter}=& -\partial_0 X\partial_0 X + 2i \left[(\chi^{(d)}_{01}+\chi^{(d)}_{10})\psi_0\partial_0 X - \chi^{(d)}_{00}\psi_1 \partial_0 X - \chi^{(d)}_{00}\psi_0 \partial_1 X\right] \\
            &\quad + 2 \psi_0\psi_1 (\chi^{(d)}_{01}+\chi^{(d)}_{10}) \chi^{(d)}_{00} + i \psi_0\partial_1\psi_0 -i \psi_1\partial_0\psi_0 -i \psi_0\partial_0\psi_1.\\
        \end{aligned}
    \end{equation}
    The SUSY transformation parameters are in doublet representation
    \begin{equation}
        \begin{aligned}
            \left[S,\begin{pmatrix} \epsilon_0\\ \epsilon_1 \end{pmatrix}\right]&=\frac{1}{2}\begin{pmatrix} 0\\ \epsilon_0 \end{pmatrix}, \\
        \end{aligned}
    \end{equation}
    and the SUSY transformation is formally the same with \eqref{eq:SUSYTransformation}. Furthermore, the super-conformal transformation is
    \begin{equation}
        \delta_{\text{SC}} \chi^{(d)}_{00} = 0 , \qquad 
        \delta_{\text{SC}} \chi^{(d)}_{01} + \delta_{\text{SC}} \chi^{(d)}_{10} = 0.
    \end{equation}
    Thus the three degrees of freedom in these two transformations just fix all components of $\chi$ field. Denoting the parameter in the super-conformal transformation as $\eta_0$, we find that the composite transformation is
    \begin{equation}
        \delta \chi^{(d)}_{00} = \nabla_0 \epsilon_0 , \qquad 
        \delta \chi^{(d)}_{01} = \nabla_0 \epsilon_1 + \eta_0, \qquad
        \delta \chi^{(d)}_{10} = \nabla_1 \epsilon_0 - \eta_0.
    \end{equation}
    Now the Faddeev-Popov determinant  \eqref{eq:FadeevPopovMeasure} can be expressed as the ghost field path-integral 
    \begin{equation}
        \begin{aligned}
            \det \left(\frac{\delta \chi^{(d)}}{\delta (\epsilon,\eta)}\right) \sim \int \mathcal{D}\beta_{00} &\mathcal{D}\beta_{01} \mathcal{D}\beta_{10} \mathcal{D}\gamma_0 \mathcal{D}\gamma_1 \delta(\beta_{01}-\beta_{10}) \\
            &\exp \left\{ -\frac{1}{2\pi} \int d^2 \sigma \left( - \gamma_1\partial_0\beta_{10} + \gamma_0\partial_0\beta_{11} -\gamma_0\partial_1\beta_{01} \right)\right\},\\
        \end{aligned}
    \end{equation}
    from which we read the Lagrangian of the $\beta\gamma$-ghost field
    \begin{equation}
        \mathcal{L}^{\text{(d-ih)}}_{\beta\gamma} =  - \gamma_1\partial_0\beta_{01} + \gamma_0\partial_0\beta_{11} -\gamma_0\partial_1\beta_{10},
    \end{equation}
    with $\beta_{01}=\beta_{10}$. Therefore we obtain the full Lagrangian of the theory, after including the one of $bc$-ghost field, 
    \begin{equation}\label{eq:InhomogeneousDoubletSuperstringLagrangian}
        \begin{aligned}
            \mathcal{L}^\text{(d-ih)}=&-\partial_0 X\partial_0 X - i c^0\partial_0 b_{00} +i c^1\partial_1b_{00} -2ic^1\partial_0 b_{01} \\
            &-i \psi_1\partial_0\psi_0 -i \psi_0\partial_0\psi_1 + i \psi_0\partial_1\psi_0  - \gamma_1\partial_0\beta_{10} + \gamma_0\partial_0\beta_{11} -\gamma_0\partial_1\beta_{10}.
        \end{aligned}
    \end{equation}  
 We will refer to this theory as the inhomogeneous doublet tensionless superstring.
 
\textbf{Vector-singlet gravitino.}
    For the gravitino in the vector-singlet representation, only $\chi^{(s)}_{01}$ contributes to the Lagrangian and the SUSY transformation. In the flat gauge, we have
    \begin{equation}
        \mathcal{L}^\text{(s-ih)}_\text{matter}=-\partial_0 X\partial_0 X + 2i \chi^{(s)}_{01}\psi_0\partial_0 X + i \psi_0\partial_1\psi_0 -i \psi_1\partial_0\psi_0 -i \psi_0\partial_0\psi_1.
    \end{equation}
    We only need one parameter $\epsilon_1$ in the singlet representation under rotation $S$ for SUSY transformation: 
    \begin{equation}
        \begin{aligned}
            &\delta_{\text{SUSY}}X=i\epsilon_1\psi_0 \qquad \delta_{\text{SUSY}} e^\alpha_a=0 \qquad \delta_{\text{SUSY}}\chi^{(s)}_{01}=\partial_0 \epsilon_1\\
            &\delta_{\text{SUSY}}\begin{pmatrix}\psi_0\\ \psi_1\end{pmatrix} = -i (\partial_0 X +i \psi_0\chi^{(s)}_{01}) \begin{pmatrix}0\\ \epsilon_1\end{pmatrix}, \qquad \text{with } \left[S,\epsilon_1\right]=0. \\
        \end{aligned}
    \end{equation}
    The singlet-SUSY transformation can be considered as turning off $\epsilon_0$ in the SUSY transformation for the vector-doublet $\chi^{(d)}$. Moreover, now there is no super-conformal transformation. Indeed, there is only one degree of freedom to fix, which can be done by the SUSY transformation. Now the transformation of $\chi^{(s)}$ is 
    \begin{equation}
        \delta \chi^{(s)}_{01}=\nabla_0\epsilon_1,
    \end{equation}
    Using
    \begin{equation}
        \det \left(\frac{\delta \chi^{(s)}_{01}}{\delta \epsilon}\right) \sim \int \mathcal{D}\beta_{01} \mathcal{D}\gamma_0 \exp \left\{ -\frac{1}{2\pi} \int d^2 \sigma - \gamma_0\partial_0\beta_{01}\right\},
    \end{equation}
    we get that the ghost Lagrangian is simply
    \begin{equation}
        \mathcal{L}^{\text{(s-ih)}}_{\beta\gamma} = -\gamma_0\partial_0\beta_{01}.
    \end{equation}
    The theory with its ghost will be referred to as the singlet theory, and the full Lagrangian is
    \begin{equation}\label{eq:InhomogeneousSingletSuperstringLagrangian}
        \begin{aligned}
            \mathcal{L}^\text{(s-ih)}=&-\partial_0 X\partial_0 X - i c^0\partial_0 b_{00} +i c^1\partial_1b_{00} -2ic^1\partial_0 b_{01} \\
            &\qquad -i \psi_1\partial_0\psi_0 -i \psi_0\partial_0\psi_1 + i \psi_0\partial_1\psi_0  -\gamma_1\partial_0\beta_{10}.
        \end{aligned}
    \end{equation}  \par

\subsection{Critical dimensions for different tensionless superstrings}\label{subsec:SuperstringCriticalDim}
    In the last sub-section we have obtained four kinds of tensionless superstring theories, the homogeneous superstring \eqref{eq:HomogeneousTensionlessSuperStringLagrangian} and its trivial reduction \eqref{eq:TrivialTensionlessSuperStringLagrangian}, the inhomogeneous doublet superstring \eqref{eq:InhomogeneousDoubletSuperstringLagrangian} and inhomogeneous singlet superstring\eqref{eq:InhomogeneousSingletSuperstringLagrangian}. In this section, we use the canonical quantization to calculate the critical dimensions of these string theories. It turns out that the canonical  quantization is suitable for the flipped vacuum $|vac\rangle_f$, which 
    is BMS invariant:
    \begin{equation}
        L_n|vac\rangle_f = M_n|vac\rangle_f = 0.
    \end{equation}
    In the following discussions, we will simply focus on the flipped vacuum, and find that 
    the critical dimensions of the homogeneous superstring and the doublet inhomogeneous superstring are the same as the usual superstring, but the ones of the trivial homogeneous superstring and the singlet inhomogeneous superstring are not integers. \par

\subsubsection*{Homogeneous and trivial tensionless superstring}
    The action of the homogeneous case in the flat gauge is
    \begin{equation}\label{eq:HomogeneousTensionlessSuperStringAction}
        \begin{aligned}
            S^\text{h}=-\frac{1}{2\pi}\int d^2\sigma \{&-\partial_0 X\partial_0 X - i c^0\partial_0 b_{00} +i c^1\partial_1b_{00} -2ic^1\partial_0 b_{01} \\
            &\quad - i\psi_0\partial_0\psi_0 - i\psi_1\partial_0\psi_1 + \gamma_0\partial_0 \beta_{0} + \gamma_1\partial_0 \beta_{1}\}.
        \end{aligned}
    \end{equation} 
    From the equations of motion, we can read the mode expansions of the fields
    \begin{equation}
    \begin{aligned}
        &X = x + \frac{1}{2}p\tau + \frac{i}{2}\sum_{n\neq0}\frac{1}{n}\left(\tilde{X}_n - in\tau X_n\right)e^{-in\sigma}, \\
        &\psi_0 = \sum_r \psi_r e^{-i r\sigma}, 
            \qquad\qquad\qquad\quad \psi_1 = \sum_r \bar{\psi}_r e^{-i r\sigma}, \\
        &b_{00} = \sum_n b_n e^{-in\sigma}, 
            \qquad\qquad\qquad\quad b_{01} = \sum_n\left(\tilde{b}_n -\frac{i}{2}n\tau b_n\right) e^{-in\sigma}, \\
        &c^0 = \sum_n\left(-\tilde{c}_n -in\tau c_n\right) e^{-in\sigma}, 
            \qquad c^1 = \sum_n c_n e^{-in\sigma}, \\
        &\beta_{0} = \sum_r \beta_r e^{-i r\sigma}, 
            \quad \beta_{1} = \sum_r \bar{\beta}_r e^{-i r\sigma}, 
            \quad \gamma_0 = \sum_r \gamma_r e^{-i r\sigma}, 
            \quad \gamma_1 = \sum_r \bar{\gamma}_r e^{-i r\sigma}.
    \end{aligned}
    \end{equation}
    The canonical quantization leads to the following nonvanishing commutation relations among the modes
    \begin{equation}
        \begin{aligned}
            &\left[x, p \right] = i, 
                &&\left[\tilde{X}_m, X_n \right] = 2m \delta_{m+n}, 
                &&\left\{\psi_r, \psi_s \right\} = \frac{1}{2}\delta_{r+s}, 
                &&\left\{\bar{\psi}_r, \bar{\psi}_s \right\} = \frac{1}{2}\delta_{r+s}, \\
            &\left\{\tilde{c}_m, b_n \right\} = -\delta_{m+n}, 
                &&\left\{\tilde{b}_m, c_n \right\} = \frac{1}{2} \delta_{m+n}, 
                &&\left[\beta_r, \gamma_s \right] = -i \delta_{r+s}, 
                &&\left[\bar{\beta}_r, \bar{\gamma}_s \right] = -i \delta_{r+s}. 
        \end{aligned}
    \end{equation}
    
    The stress tensors give the  conserved Noether current of the diffeomorphism. After using the equations of motions, they have the following forms
    \begin{equation}
        \begin{aligned}
        &\begin{aligned}
            &{T_{X}}^\alpha_{~\beta}= \frac{1}{2\pi} \begin{pmatrix} 
                    \partial_0 X \partial_0 X & 2\partial_0 X \partial_1 X \\ 
                    0 & - \partial_0 X \partial_0 X \\
                \end{pmatrix}, \\
        \end{aligned}\\[15pt]
        &\begin{aligned}
            &{T_{bc}}^\alpha_{~\beta}= \frac{1}{2\pi} \begin{pmatrix} 
                    -i c^1 \partial_1 b_{00} -2i \partial_1 c^1 b_{00} & -i c^0 \partial_1 b_{00} -2i \partial_1 c^0 b_{00} -2i c^1 \partial_1 b_{01} -4i \partial_1 c^1 b_{01} \\ 
                    0 & i c^1 \partial_1 b_{00} +2i \partial_1 c^1 b_{00}  \\
                \end{pmatrix},  \\
        \end{aligned}\\[15pt]
        &\begin{aligned}
            &{T_{\psi}}^\alpha_{~\beta}= \frac{1}{2\pi} \begin{pmatrix} 
                    0 & i\psi_0 \partial_1 \psi_0 +i\psi_1 \partial_1 \psi_1 \\ 
                    0 & 0 \\
                \end{pmatrix},\\
        \end{aligned}\\[15pt]
        &\begin{aligned}
            &{T_{\beta\gamma}}^\alpha_{~\beta}= \frac{1}{2\pi} \begin{pmatrix} 
                    0 & \frac{3}{2}\partial_1 \gamma_0 \beta_0 +\frac{1}{2}\gamma_0 \partial_1 \beta_0 +\frac{3}{2}\partial_1 \gamma_1 \beta_1 +\frac{1}{2}\gamma_1 \partial_1 \beta_1 \\ 
                    0 & 0 \\
                \end{pmatrix}. \\
        \end{aligned}
        \end{aligned}
    \end{equation}
    Thus the corresponding conserved charges are
    \begin{equation}
        \begin{aligned}
            L_n &=\sum_m \frac{1}{2} :X_{n-m}\tilde{X}_m: + (m-2n) \left(-:\tilde{c}_{n-m} b_{m}: +2 :c_{n-m} \tilde{b}_{m}: \right) \\
                &\qquad + \sum_r \left(r-\frac{1}{2}n\right) \left(:\psi_{n-r}\psi_r: +:\bar{\psi}_{n-r} \bar{\psi}_r: \right) + i \left( r -\frac{3}{2}n \right) \left(:\gamma_{n-r}\beta_r: +:\bar{\gamma}_{n-r} \bar{\beta}_r: \right),\\
            M_n &=\sum_m \frac{1}{4} X_{n-m} X_m + (m-2n) ~ c_{n-m} b_{m}, \\
        \end{aligned}
    \end{equation}
    which generate the BMS symmetries. 
    
    The SUSY transformation on the fields take the following forms
    \begin{equation}
        \begin{aligned}
            &\delta X= i\epsilon_1\psi_0 -i\epsilon_0\psi_1, 
                \quad \delta \psi_0 = -\epsilon_1 \partial_0 X,
                \quad \delta \psi_1 = \epsilon_0 \partial_0 X, \\
            &\delta b_{00} = -\frac{1}{2}\epsilon_1\partial_0\beta_{0} +\frac{1}{2}\epsilon_0\partial_0\beta_{1}, 
                \quad \delta b_{01} = -\frac{1}{2}\epsilon_1\partial_1\beta_{0} +\frac{1}{2}\epsilon_0\partial_1\beta_{1}, \\
            &\delta c^0 = \epsilon_1\gamma_0 -\epsilon_0\gamma_1, 
                \qquad\qquad\quad~~ \delta c^1 = 0, \\
            &\delta \beta_{0} = i\epsilon_1 b_{00}, 
                \qquad\qquad\qquad\qquad \delta \beta_{1} = -i\epsilon_0 b_{00}, \\
            &\delta \gamma_0 = -\frac{1}{2}i\epsilon_1\partial_0 c^0 -\frac{1}{2}i\epsilon_1\partial_1 c^1, 
                \quad \delta \gamma_1 = \frac{1}{2}i\epsilon_0\partial_0 c^0 +\frac{1}{2}i\epsilon_0\partial_1 c^1. \\
        \end{aligned}
    \end{equation}
    The corresponding conserved supercurrents are
    \begin{equation}
        \begin{aligned}
            &(J_{X\psi})^\alpha_{A=0}= \frac{1}{2\pi} \begin{pmatrix} 2\psi_0\partial_0 X \\ 0 \end{pmatrix}, \qquad \qquad 
            (J_{X\psi})^\alpha_{A=1}= \frac{1}{2\pi} \begin{pmatrix} 2\psi_1\partial_0 X \\ 0 \end{pmatrix}, \\
            &(J_{\text{ghost}})^\alpha_{A=0}= \frac{1}{2\pi} \begin{pmatrix}
                -c^1\partial_1\beta_{0} +\gamma_0 b_{00} + 4\partial_1\left(c^1\beta_{0}\right) \\ 0 \end{pmatrix}, \\
            &(J_{\text{ghost}})^\alpha_{A=1}= \frac{1}{2\pi} \begin{pmatrix}
                -c^1\partial_1\beta_{1} +\gamma_1 b_{00} + 4\partial_1\left(c^1\beta_{1}\right) \\ 0 \end{pmatrix}. \\
        \end{aligned}
    \end{equation}
    The conserved charges are defined as
    \begin{equation}
        \begin{aligned}
            H^{(0)}_r &=\int d\sigma~ J^0_{~0} e^{i r\sigma},
            & H^{(1)}_r &=\int d\sigma~ J^0_{~1} e^{i r\sigma},\\
        \end{aligned}
    \end{equation}
    which can be expressed in terms of the modes
    \begin{equation}
        \begin{aligned}
            H^{(0)}_r &=\sum_m \psi_{r-m} X_m -i (3r+m) c_m \beta_{r-m} +\gamma_{r-m} b_m,\\
            H^{(1)}_r &=\sum_m \bar{\psi}_{r-m} X_m -i (3r+m) c_m \bar{\beta}_{r-m} +\bar{\gamma}_{r-m} b_m.\\
        \end{aligned}
    \end{equation}
    Together with the BMS algebra, they form the homogenenous super-BMS algebra with the commutation relations
    \begin{equation}
        \begin{aligned}
            &\left[L_m, L_n \right] = \left(m-n\right) L_{m+n} + \delta_{m+n}A_L(m), \qquad &&\left[L_m, M_n\right] = \left(m-n\right) M_{m+n}+ \delta_{m+n}A_M(m),\\
            &\left[L_n, H^{(i)}_r \right] = \left(\frac{1}{2}n-r\right) H^{(i)}_{n+r}, \qquad &&\left[M_n, H^{(i)}_r\right] = 0,\\
            &\left\{ H^{(i)}_r, H^{(j)}_s\right\} = 2\delta^{ij} M_{r+s},
        \end{aligned}
    \end{equation}
    where $A_L(m)$ and $A_M(m)$ are anomaly terms. Noticing that there is no anomaly term in the (anti-)commutation relations involving the fermionic generators. Using the Jacobi identity of $\{L_m,L_n,L_k\}$ and $\{L_m,L_n,M_k\}$, we can fix $A_L(m)$ and $A_M(m)$ to be 
    \begin{equation}
        \begin{aligned}
            &A_L(m)=c^L_3 m^3 + c^L_1 m, && A_M(m)=c^M_3 m^3 + c^M_1 m. \\
        \end{aligned}
    \end{equation}
    Considering the following two expectation values in the flipped vacuum
    \begin{equation}\nonumber
        \begin{aligned}
            _f\langle vac|[L_1,M_{-1}]|vac\rangle_f &= \langle 2 M_0\rangle_f +A_M(1), &_f\langle vac|[L_2,M_{-2}]|vac\rangle_f &= \langle 4 M_0\rangle_f +A_M(2), 
        \end{aligned}
    \end{equation}
    we find
    \begin{equation}
        \begin{aligned}
            &A_M(m)=0. \\
        \end{aligned}
    \end{equation}
    And from the expectation values
    \begin{equation}\nonumber
        \begin{aligned}
            _f\langle vac|[L_1,L_{-1}]|vac\rangle_f &= \langle 2 L_0\rangle_f + A_L(1), &_f\langle vac|[L_2,L_{-2}]|vac\rangle_f &= \langle 4 L_0\rangle_f +A_L(2), \\
        \end{aligned}
    \end{equation}
    we get
    \begin{equation}
        \begin{aligned}
            &&A^X_L(m)&=\frac{D}{6}(m^3-m), & A^{bc}_L(m)&=-\frac{1}{3}(13m^3-m), \\
            \text{NS sector:} && A^\psi_L(m)&=\frac{D}{12}(m^3-m), & A^{\beta\gamma}_L(m)&=\frac{1}{6}(11m^3+m), \\
            \text{R sector:} && A^\psi_L(m)&=\frac{D}{12}(m^3+2m), & A^{\beta\gamma}_L(m)&=\frac{1}{6}(11m^3-2m). \\
        \end{aligned}
    \end{equation}
    As the total anomaly $A_L(m)=A^X_L(m)+A^{bc}_L(m)+A^\psi_L(m)+A^{\beta\gamma}_L(m)+2a_L m$ should be vanishing, we can fix the critical dimension and the normal ordering constant $a$ in the  tensionless homogeneous superstring
    \begin{equation}\label{eq:N=2HomogeneousCriticalDimension}
        \begin{aligned}
            \text{NS sector:} && D=10, \qquad a_L=1, \\
            \text{R sector:} && D=10, \qquad a_L=0. \\
        \end{aligned}
    \end{equation}
    
    As we showed before, the above action can be further simplified by setting half of the fermionic fields vanishing. The resulting tensionless trivial superstring has the following action in the flat gauge:
    \begin{equation}\label{eq:TrivialTensionlessSuperStringAction}
        \begin{aligned}
            S^\text{tri}=-\frac{1}{2\pi}\int d^2\sigma \{&-\partial_0 X\partial_0 X - i c^0\partial_0 b_{00} +i c^1\partial_1b_{00} -2ic^1\partial_0 b_{01} - i\psi_0\partial_0\psi_0 + \gamma_0\partial_0 \beta_{0} \}.
        \end{aligned}
    \end{equation}
    Compared with the action \eqref{eq:TrivialTensionlessSuperStringAction}, the action \eqref{eq:HomogeneousTensionlessSuperStringAction} could be viewed as the one for a $\mathcal{N}=2$ superstring theory, while the trivial one could be taken as an $\mathcal{N}=1$ theory. One may construct the theories with more supersymmetries as well. In all these cases, the discussion is similar and straightforward. However, in the  case of trivial tensionless superstring, the critical dimension now is not an integer, which implies this superstring theory is not a good one. In fact, the total anomalies of NS- and R-sectors in the theories with $\mathcal{N}$ supersymmetries are, respectively, 
    \begin{equation}
        \begin{aligned}
            A^{\text{NS}}_L(m) &=\frac{D}{6}(m^3-m) -\frac{1}{3}(13m^3-m) + \frac{D\mathcal{N}}{24}(m^3-m)+\frac{\mathcal{N}}{12}(11m^3+m)+2a_L m, \\
            A^{\text{R}}_L(m) &=\frac{D}{6}(m^3-m) -\frac{1}{3}(13m^3-m) + \frac{D\mathcal{N}}{24}(m^3+2m)+\frac{\mathcal{N}}{12}(11m^3-2m)+2a_L m. \\
        \end{aligned}
    \end{equation}
    Requiring the vanishing of the anomalies $A_L(m)$ gives the critical dimensions and the normal-ordering constants. The cases with positive critical dimensions are listed in Table \ref{tb:CriticalDimForHomogeneous}, and especially, the $\mathcal{N}=2$ case corresponds to \eqref{eq:N=2HomogeneousCriticalDimension}. We see that only the theories with $\mathcal{N}=2$ or $\mathcal{N}=4$ have integer-valued critical dimensions. The theories with $\mathcal{N}\ge 5$ have negative critical dimensions. 
    
    \begin{table}[ht]
        \def\arraystretch{1.6}
        \centering
        \caption{\centering Critical dimensions of homogeneous tensionless superstring theories}
        \label{tb:CriticalDimForHomogeneous}
        \begin{tabular}{cccc}
            \hline
            boundary condition & $\mathcal{N}$  & critical dimension & normal-ordering constant\\
            \hline
            \multirow{4}{*}{NS} & $1$ & $\frac{82}{5}$ & $\frac{3}{2}$\\
            ~ & $2$ & $10$ & $1$\\
            ~ & $3$ & $\frac{38}{7}$ & $\frac{1}{2}$\\
            ~ & $4$ & $2$ & $0$\\
            \hline
            \multirow{4}{*}{R} & $1$ & $\frac{82}{5}$ & $\frac{3}{5}$\\
            ~ & $2$ & $10$ & $0$\\
            ~ & $3$ & $\frac{38}{7}$ & $-\frac{1}{7}$\\
            ~ & $4$ & $2$ & $0$\\
            \hline
        \end{tabular}
    \end{table}\par
    
\subsubsection*{Inhomogeneous doublet tensionless superstring}
    The action of the inhomogeneous doublet superstring theory in the flat gauge is
    \begin{equation}\label{eq:InhomogeneousDoubletSuperstringAction}
        \begin{aligned}
            S^\text{d-ih}=&-\frac{1}{2\pi}\int d^2\sigma \{-\partial_0 X\partial_0 X - i c^0\partial_0 b_{00} +i c^1\partial_1b_{00} -2ic^1\partial_0 b_{01} \\
            &-i \psi_1\partial_0\psi_0 -i \psi_0\partial_0\psi_1 + i \psi_0\partial_1\psi_0  - \gamma_1\partial_0\beta_{10} + \gamma_0\partial_0\beta_{11} -\gamma_0\partial_1\beta_{10}\}.
        \end{aligned}
    \end{equation}
    The mode expansions of the fields are
    \begin{equation}
        \begin{aligned}
            &X = x + \frac{1}{2}p\tau + \frac{i}{2}\sum_{n\neq0}\frac{1}{n}\left(\tilde{X}_n - in\tau X_n\right)e^{-in\sigma}, \\
            &\psi_0 = \sum_r \psi_r e^{-i r\sigma}, 
                \qquad\qquad\qquad\qquad \psi_1 = \sum_r \left(\frac{1}{2}\tilde{\psi}_r -i r\tau \psi_n\right) e^{-i r\sigma}, \\
            & b_{00} = \sum_n b_n e^{-in\sigma},  
                \qquad\qquad\qquad\qquad b_{01} = \sum_n\left(\tilde{b}_n -\frac{i}{2}n\tau b_n\right) e^{-in\sigma}, \\
            &c^0 = \sum_n\left(-\tilde{c}_n -in\tau c_n\right) e^{-in\sigma}, 
                ~~~\qquad c^1 = \sum_n c_n e^{-in\sigma}, \\
            &\beta_{10} = \sum_r \beta_r e^{-i r\sigma}, 
                \qquad\qquad\qquad\qquad \beta_{11} = \sum_r \left(\frac{3}{2}\tilde{\beta}_r -i r\tau \beta_r\right) e^{-ir\sigma}, \\
            &\gamma_0 = \sum_r \gamma_r e^{-i r\sigma}, 
                ~~\qquad\qquad\qquad\qquad \gamma_1 = \sum_r\left(\frac{1}{2}\tilde{\gamma}_r +i r\tau \gamma_r\right) e^{-ir\sigma}, \\
        \end{aligned}
    \end{equation}
    with the commutation relations among the modes
    \begin{equation}
        \begin{aligned}
            &\left[x, p \right] = i, &&\left[\tilde{X}_m, X_n \right] = 2m \delta_{m+n}, &&\left\{\tilde{\psi}_r, \psi_s \right\} = \delta_{r+s}, \\
            &\left\{\tilde{c}_m, b_n \right\} = -\delta_{m+n}, &&\left\{\tilde{b}_m, c_n \right\} = \frac{1}{2} \delta_{m+n}, &&\left[\tilde{\gamma}_r, \beta_s \right] = -2i \delta_{r+s}, &&\left[\gamma_r, \tilde{\beta}_s \right] = \frac{2}{3} i \delta_{r+s}. \\
        \end{aligned}
    \end{equation}

   Similar to the homogeneous case, we can read the conserved charges from the Noether currents, corresponding to the diffeomorphism and supersymmetric transformations. In the end, we find    
    \begin{equation}
        \begin{aligned}
            L_n &=\sum_m \frac{1}{2}:X_{n-m}\tilde{X}_m: + (m-2n) \left(-:\tilde{c}_{n-m} b_{m}: +2 :c_{n-m} \tilde{b}_{m}: \right)  \\
                &\qquad + \sum_r \left(r-\frac{1}{2}n\right) :\psi_{n-r}\tilde{\psi}_r: + i\left( r-\frac{3}{2}n \right) \left(\frac{3}{2}:\gamma_{n-r}\tilde{\beta}_{r}: -\frac{1}{2}:\tilde{\gamma}_{n-r}\beta_{r}:\right),\\
            M_n &=\sum_m \frac{1}{4} X_{n-m} X_m + (m-2n) ~ c_{n-m} b_{m} \\
                &\qquad\qquad\qquad + \sum_r \left(r-\frac{1}{2}n\right)~\psi_{n-r}\psi_r + i\left( r-\frac{3}{2}n \right) \gamma_{n-r}\beta_r.\\
            G_r &=\sum_m \psi_{r-m}\tilde{X}_m +\frac{1}{2}\tilde{\psi}_{r-m} X_m \\
                &\qquad\qquad -i (2r+m) \left(-\tilde{c}_m\beta_{r-m} +\frac{3}{2} c_m\tilde{\beta}_{r-m}\right) -\frac{1}{2}\tilde{\gamma}_{r-m} b_m +2\gamma_{r-m} \tilde{b}_m,\\
            H_r &=\sum_m \psi_{r-m} X_m -i (2r+m) c_m\beta_{r-m} +\gamma_{r-m} b_m.\\
        \end{aligned}
    \end{equation}
    which generate the inhomogenenous doublet super-BMS algebra with the commuation relations
    \begin{equation}\label{eq:InhomoDoubletSUSYAlgebra}
        \begin{aligned}
            &\left[L_m, L_n \right] = \left(m-n\right) L_{m+n} + \delta_{m+n}A_L(m), \qquad &&\left[L_m, M_n\right] = \left(m-n\right) M_{m+n} + \delta_{m+n}A_M(m),\\
            &\left[L_n, G_r \right] = \left(\frac{1}{2}n-r\right) G_{n+r}, \qquad &&\left[M_n, G_r\right] = \left(\frac{1}{2}n-r\right) H_{n+r},\\
            &\left[L_n, H_r \right] = \left(\frac{1}{2}n-r\right) H_{n+r}, \qquad &&\left[M_n, H_r\right] = 0,\\
            &\left\{G_r, G_s\right\} = 2 L_{r+s} + \delta_{r+s}B_G(r), \qquad &&\left\{G_r, H_s\right\} = 2 M_{r+s}, \\
            &\left\{H_r, H_s\right\} = 0.&&
        \end{aligned}
    \end{equation}
    where $A_L(m)$, $A_M(m)$ and $B_G(r)$ are all anomaly terms. Using the Jacobi identity of $\{L_m,L_n,L_k\}$, $\{L_m,L_n,M_k\}$ and $\{L_m,G_r,G_s\}$, we can fix the anomaly terms to be 
    \begin{equation}
        \begin{aligned}
            &A_L(m)=c^L_3 m^3 + c^L_1 m, \\
            &A_M(m)=c^M_3 m^3 + c^M_1 m, \\
            &B_G(r)=\left(4A_L(1)+\frac{4}{3}c^B\right) r^2 - \frac{1}{3}c^B. \\
        \end{aligned}
    \end{equation}
    Similar to the homogeneous case, considering the expectation values
    \begin{equation}\nonumber
        \begin{aligned}
            _f\langle vac|[L_1,M_{-1}]|vac\rangle_f &= \langle 2 M_0\rangle_f +A_M(1), &_f\langle vac|[L_2,M_{-2}]|vac\rangle_f &= \langle 4 M_0\rangle_f +A_M(2), \\
        \end{aligned}
    \end{equation}
    we get
    \begin{equation}
        \begin{aligned}
            &A_M(m)=0. \\
        \end{aligned}
    \end{equation}
    And from the expectation values
    \begin{equation}\nonumber
        \begin{aligned}
            _f\langle vac|[L_1,L_{-1}]|vac\rangle_f &= \langle 2 L_0\rangle_f + A_L(1), &_f\langle vac|[L_2,L_{-2}]|vac\rangle_f &= \langle 4 L_0\rangle_f +A_L(2), \\
        \end{aligned}
    \end{equation}
    we find
    \begin{equation}\label{eq:InhomoDoubletAnomalyInyPart}
        \begin{aligned}
            &&A^X_L(m)&=\frac{D}{6}(m^3-m), & A^{bc}_L(m)&=-\frac{1}{3}(13m^3-m), \\
            \text{NS sector:} && A^\psi_L(m)&=\frac{D}{12}(m^3-m), & A^{\beta\gamma}_L(m)&=\frac{1}{6}(11m^3+m), \\
            \text{R sector:} && A^\psi_L(m)&=\frac{D}{12}(m^3+2m), & A^{\beta\gamma}_L(m)&=\frac{1}{6}(11m^3-2m), \\
        \end{aligned}
    \end{equation}
    The anomaly cancellation helps us to determine  the critical dimension and the normal ordering constant $a$,
    \begin{equation}\label{eq:CriticalDimOfInhomoDoublet}
        \begin{aligned}
            \text{NS sector:} && D=10, \qquad a_L=1, \\
            \text{R sector:} && D=10, \qquad a_L=0. \\
        \end{aligned}
    \end{equation}
    Moreover, there is still  the anomaly term $B_G(r)=B^{X\psi}_G(r)+B^{\text{ghost}}_G(r)+2a_L $. From the expectation values
    \begin{equation}\nonumber
        \begin{aligned}
            \text{NS sector:} && _f\langle vac|[G_{\frac{1}{2}},G_{-\frac{1}{2}}]|vac\rangle_f &= \langle 2 L_0\rangle_f + B_G\left(\frac{1}{2}\right), \\
                &&_f\langle vac|[G_{\frac{3}{2}},G_{-\frac{3}{2}}]|vac\rangle_f &= \langle 2 L_0\rangle_f +B_G\left(\frac{3}{2}\right), \\[10pt]
            \text{R sector:} && _f\langle vac|[G_{0},G_{0}]|vac\rangle_f &= \langle 2 L_0\rangle_f + B_G(0), \\
                &&_f\langle vac|[G_{1},G_{-1}]|vac\rangle_f &= \langle 2 L_0\rangle_f +B_G(1), \\
        \end{aligned}
    \end{equation}
    we have
    \begin{equation}
        \begin{aligned}
            \text{NS sector:} && B^{X\psi}_G(r)&=\frac{D}{2}r^2 - \frac{D}{8}, &B^{\text{ghost}}_G(r)&=-5r^2 - \frac{3}{4},\\
            \text{R sector:} && B^{X\psi}_G(r)&=Dr^2, &B^{\text{ghost}}_G(r)&=-10r^2.\\
        \end{aligned}
    \end{equation}
   The anomaly cancellation in $B_G(r)$  leads to  the same result as \eqref{eq:CriticalDimOfInhomoDoublet}. \par

    For the theories with extended supersymmetries, the algebra are different from \eqref{eq:InhomoDoubletSUSYAlgebra}. Nevertheless, we can fix the critical dimensions by cancelling the anomaly $A_L(m)$. Using the result in \eqref{eq:InhomoDoubletAnomalyInyPart}, we find the positive critical dimensions and the corresponding normal-ordering constants, which are listed in Table \ref{tb:CriticalDimForInhomoDoublet}. 
    
    \begin{table}[ht]
        \def\arraystretch{1.6}
        \centering
        \caption{\centering Critical dimensions of inhomogeneous doublet tensionless superstring theories}
        \label{tb:CriticalDimForInhomoDoublet}
        \begin{tabular}{cccc}
            \hline
            boundary condition & $\mathcal{N}$  & critical dimension & normal-ordering constant\\
            \hline
            \multirow{2}{*}{NS} & $1$ & $10$ & $1$\\
            ~ & $2$ & $2$ & $0$\\
            \hline
            \multirow{2}{*}{R} & $1$ & $10$ & $0$\\
            ~ & $2$ & $2$ & $0$\\
            \hline
        \end{tabular}
    \end{table}\par
    
\subsubsection*{Inhomogeneous singlet tensionless superstring}
    The action of the inhomogeneous singlet superstring in the flat gauge is
    \begin{equation}\label{eq:InhomogeneousSingletSuperstringAction}
        \begin{aligned}
            S^\text{s-ih}=-\frac{1}{2\pi}\int d^2\sigma \{&-\partial_0 X\partial_0 X - i c^0\partial_0 b_{00} +i c^1\partial_1b_{00} -2ic^1\partial_0 b_{01} \\
            &-i \psi_1\partial_0\psi_0 -i \psi_0\partial_0\psi_1 + i \psi_0\partial_1\psi_0  -\gamma_1\partial_0\beta_{10}\}.
        \end{aligned}
    \end{equation}
    Obviously, only the $\beta\gamma$ part is different from the one in the doublet theory, and we only need to study this part. The mode expansions are
    \begin{equation}
        \begin{aligned}
            &\begin{aligned}
                &\beta_{10} = \sum_r \beta_r e^{-i r\sigma}, \\
            \end{aligned} \quad
            &&\begin{aligned}
                &\gamma_1 = \sum_r \tilde{\gamma}_r e^{-i r\sigma}, \\
            \end{aligned}\\
        \end{aligned}
    \end{equation}
    with the commutation relation
    \begin{equation}
        \begin{aligned}
            \begin{aligned}
                &\left[\beta_r, \tilde{\gamma}_s \right] = i \delta_{r+s}. \\
            \end{aligned}
        \end{aligned}
    \end{equation}
   
    The conserved charges generating the BMS algebra are
    \begin{equation}
        \begin{aligned}
            L_n &=\sum_m \frac{1}{2}:X_{n-m}\tilde{X}_m: + (m-2n) \left(-:\tilde{c}_{n-m} b_{m}: +2 :c_{n-m} \tilde{b}_{m}: \right)  \\
                &\qquad + \sum_r \left(r-\frac{1}{2}n\right) :\psi_{n-r}\tilde{\psi}_r: - i\left( r-\frac{3}{2}n \right):\tilde{\gamma}_{n-r}\beta_{r}:,\\
            M_n &=\sum_m \frac{1}{4} X_{n-m} X_m + (m-2n) ~ c_{n-m} b_{m}  + \sum_r \left(r-\frac{1}{2}n\right)~\psi_{n-r}\psi_r. \\
        \end{aligned}
    \end{equation}
    The supercurrent gives only one conserved charge
     \begin{equation}
        \begin{aligned}
            H_r &=\sum_m \psi_{r-m} X_m +\tilde{\gamma}_{r-m} b_m.
        \end{aligned}
    \end{equation}
    Together with the generators of the BMS algebra, they generate the inhomogenenous singlet super-BMS algebra:
    \begin{equation}
        \begin{aligned}
            &\left[L_m, L_n \right] = \left(m-n\right) L_{m+n}, \qquad \left[L_m, M_n\right] = \left(m-n\right) M_{m+n},\\
            &\left[L_n, H_r \right] = \left(\frac{1}{2}n-r\right) H_{n+r}, \qquad \left[M_n, H_r\right] = 0, \qquad \left\{H_r, H_s\right\} = 0,&\\
        \end{aligned}
    \end{equation}
    where $A_L(m)$ and $A_M(m)$ are anomaly terms.  Using the Jacobi identity of $\{L_m,L_n,L_k\}$ and $\{L_m,L_n,M_k\}$, we can fix $A_L(m)$ and $A_M(m)$ 
    \begin{equation}
        \begin{aligned}
            &A_L(m)=c^L_3 m^3 + c^L_1 m, &&A_M(m)=c^M_3 m^3 + c^M_1 m. \\
        \end{aligned}
    \end{equation}
    Similar to the discussions above, we can find 
    \begin{equation}
        \begin{aligned}
            &A_M(m)=0, \\
        \end{aligned}
    \end{equation}
    and 
    \begin{equation}
        \begin{aligned}
            &&A^X_L(m)&=\frac{D}{6}(m^3-m), & A^{bc}_L(m)&=-\frac{1}{3}(13m^3-m), \\
            \text{NS sector:} && A^\psi_L(m)&=\frac{D}{12}(m^3-m), & A^{\beta\gamma}_L(m)&=\frac{1}{12}(11m^3+m), \\
            \text{R sector:} && A^\psi_L(m)&=\frac{D}{12}(m^3+2m), & A^{\beta\gamma}_L(m)&=\frac{1}{12}(11m^3-2m), \\
        \end{aligned}
    \end{equation}
    The anomaly cancellation furthermore requires that 
    \begin{equation}
        \begin{aligned}
            \text{NS sector:} && D&=\frac{41}{3}, &a_L &=\frac{3}{2}, \\
            \text{R sector:} && D&=\frac{41}{3}, &a_L &=-\frac{1}{12}. \\
        \end{aligned}
    \end{equation}
    For the theories with general extended supersymmetries, the requirement of cancelling the anomaly results in fractional critical dimensions. The theories of positive critical dimensions require $\mathcal{N}\le 4$ and are listed in Table \ref{tb:CriticalDimForInhomoSinglet}.

    \begin{table}[ht]
        \def\arraystretch{1.6}
        \centering
        \caption{\centering Critical dimensions of inhomogeneous singlet tensionless superstring theories}
        \label{tb:CriticalDimForInhomoSinglet}
        \begin{tabular}{cccc}
            \hline
            boundary condition & $\mathcal{N}$  & critical dimension & normal-ordering constant\\
            \hline
            \multirow{4}{*}{NS} & $1$ & $\frac{41}{3}$ & $\frac{3}{2}$\\
            ~ & $2$ & $\frac{15}{2}$ & $1$\\
            ~ & $3$ & $\frac{19}{5}$ & $\frac{1}{2}$\\
            ~ & $4$ & $\frac{4}{3}$ & $0$\\
            \hline
            \multirow{4}{*}{R} & $1$ & $\frac{41}{3}$ & $-\frac{1}{12}$\\
            ~ & $2$ & $\frac{15}{2}$ & $-\frac{5}{8}$\\
            ~ & $3$ & $\frac{19}{5}$ & $-\frac{11}{20}$\\
            ~ & $4$ & $\frac{4}{3}$ & $-\frac{1}{6}$\\
            \hline
        \end{tabular}
    \end{table}\par



\section{BMS free theories: \texorpdfstring{$bc$}~ and \texorpdfstring{$\beta\gamma$}~ ghosts} \label{multiplet}

As we shown in  previous sections,  in the path integral formulation of the tensionless (super)string, we need to introduce   the BMS $bc$ ghosts and  $\beta\gamma$ ghosts to account for the Faddeev-Popov determinants. These two kinds of ghosts can be obtained by taking the inhomogenous UR limits of the usual ghosts as well. In this section, we are going to study the properties of BMS $bc$ and $\beta\gamma$ field theories in more details.

In fact, the BMS $bc$ and $\beta\gamma$ field theories  present new kinds of  BMS free field theories. The other  BMS free theories, the BMS free scalar and free fermions, had been studied in the literatures \cite{Hao:2021urq,Yu:2022bcp,Hao:2022xhq,Banerjee:2022ocj}. These theories exhibit novel features such as  the staggered modules and enhanced underlying symmetries, besides the expected appearance of the boost multiplets\cite{Chen:2020vvn}. Here we aim to investigate if these novel features persist in the BMS $bc$ and $\beta\gamma$ field theories as well. 
 
 In the BMS free fermions, it has been shown that the underlying symmetry algebra is a  BMS-Kac-Moody algebra with a $U(1)$ Kac-Moody subalgebra \cite{Yu:2022bcp}. As we will show in the following, the underlying symmetry of the  BMS $bc$ ghosts is also generated by a BMS-Kac-Moody algebra, but now the Kac-Moody subalgebra is built from  a three-dimensional  non-abelian and non-semi-simple Lie algebra.  Besides, We will  show that the symmetry algebra of the BMS free scalar also contains a non-abelian and non-semisimple  Kac-Moody subalgebra.
 
 Unlike the discussion in  previous sections,  here we will study the ghost theories defined on the plane, whose coordinates are denoted as $x$ and $y$.  Our discussion  will be sightly more general to include BMS ghosts with general scaling dimensions.
 Because the discussion of the BMS $\beta\gamma$ ghosts is quite similar with the BMS $bc$ ghosts, we will mainly focus on the BMS $bc$ ghosts. 
 


We start with the action of the BMS $bc$ ghosts
\begin{equation}
    S=\frac{1}{2\pi}\int d^2\sigma (c^0\partial_0b^0-c^1\partial_1b^0+2c^1\partial_0b^1).
\end{equation}
In this theory, the fundamental fields $\mathbf{b}\equiv(b^1, b^0)^{\mathrm{T}}$ and $\mathbf{c}\equiv(c^0, c^1)^{\mathrm{T}}$ form two rank-$2$ boost multiplets. Their conformal dimensions and boost charges are, respectively,
\begin{equation} \label{Delta and xi}
\mathbf{\Delta}_{\mathbf{b}}=\left(\begin{matrix}
   2 &0  \\
   0& 2\end{matrix}\right),
  \boldsymbol{\xi}_{\mathbf{b}}=\left(\begin{matrix}
   0&0  \\
   1&0\end{matrix}\right), \qquad
   \mathbf{\Delta}_{\mathbf{c}}=\left(\begin{matrix}
   -1 &0  \\
   0& -1\end{matrix}\right),
  \boldsymbol{\xi}_{\mathbf{c}}=\left(\begin{matrix}
   0&0  \\
   -1&0\end{matrix}\right).
\end{equation}
This can be easily seen from the infinitesimal transformation law of the field $b^1, b^0, c^0, c^1$ in \eqref{trans law}. Recall that a rank-$r$ multiplet has the following behaviour under a finite BMS transformation \cite{Chen:2020vvn,Hao:2021urq}:
\begin{equation}
    \tilde{O}_a(\tilde{\sigma}, \tilde{\tau})=\sum_{k=0}^{a}\frac{1}{k!}|f'|^{-\Delta}\partial_\xi^ke^{-\xi\frac{g'+f''}{f'}}O_{a-k}(\sigma,\tau)  \label{multiplet trans}
\end{equation}
where $O_{a}(x,y), a=0,1,...,r-1$ is the $a$-th primary field in the multiplet, and
\begin{equation}
    \tilde{\sigma}=f(\sigma), \quad \tilde{\tau}=f'(\sigma)\tau+g(\sigma)
\end{equation}
is a finite BMS transformation. From the transformation law \eqref{multiplet trans}, one can get the infinitesimal form of the transformation. In particular, for the rank-2 case, the transformations turn out to be
\begin{equation}
\begin{aligned}
    \delta_{\epsilon,\omega} O_{0}(\sigma,\tau)&=(\Delta \epsilon'+\epsilon\partial_\sigma+\omega\partial_\tau)O_{0}(\sigma,\tau),\\
    \delta_{\epsilon,\omega} O_{1}(\sigma,\tau)&=(\Delta \epsilon'+\epsilon\partial_\sigma+\omega\partial_\tau)O_{1}(\sigma,\tau)+\partial_\sigma\omega O_{0}(\sigma,\tau),
\end{aligned}
\end{equation}
where $\epsilon\equiv\epsilon(\sigma)$ and $\omega\equiv\omega(\sigma,\tau)=\epsilon'(\sigma)\tau+\upsilon(\sigma)$ are some infinitesimal transformations. Identifying $\epsilon, \omega$ with $\xi^1, \xi^0$ in Eq. \eqref{trans law} and noticing the identity $\partial_\tau\omega=\epsilon'$, we find that $\mathbf{b}, \mathbf{c}$ are both rank-2 multiplets with quantum numbers \eqref{Delta and xi}.

As usually do in studying 2D conformal field theory, we would like to do the analysis on the plane. Through the plane-cylinder map
\begin{equation}
    x=e^{i\sigma}, \quad y=i\tau e^{i\sigma},
\end{equation}
and making use of the transformation law \eqref{multiplet trans} for the multiplets  $\mathbf{b}, \mathbf{c}$ (here $f(\sigma)=e^{i\sigma}, g(\sigma)=0$), we find the action of $bc$ ghosts on the plane has a similar form with the one on the cylinder:
\begin{equation}\label{actionplane}
     S=\frac{1}{2\pi}\int dxdy (c^0\partial_y b^0-c^1\partial_x b^0+2c^1\partial_y b^1).
\end{equation}
Besides, we  want to study more general BMS $bc$-ghosts, whose conformal dimensions and boost charges are
\begin{equation} \label{general case}
\mathbf{\Delta}_{\mathbf{b}}=\left(\begin{matrix}
   \eta &0  \\
   0& \eta\end{matrix}\right),
  \boldsymbol{\xi}_{\mathbf{b}}=\left(\begin{matrix}
   0&0  \\
   \frac{\eta}{2}&0\end{matrix}\right), \qquad
   \mathbf{\Delta}_{\mathbf{c}}=\left(\begin{matrix}
   1-\eta &0  \\
   0& 1-\eta\end{matrix}\right),
  \boldsymbol{\xi}_{\mathbf{c}}=\left(\begin{matrix}
   0&0  \\
   1-\eta&0\end{matrix}\right).
\end{equation}
 Note that the above quantum numbers still keep the invariance of the action \eqref{actionplane} under the BMS transformations, and  when $\eta=2$ they return to the ones of the BMS $bc$-ghosts arising from the path-integral of the tensionless string.
 
One can easily work out  the canonical quantization, just as what we have done on the cylinder (and for $\eta=2$). The mode expansion now becomes:
\begin{equation}
\begin{aligned}
     b^0(x,y)&=\sum_n \tilde{b}_n x^{-n-\eta}, \quad b^1(x,y)=\frac{1}{2}\sum_n [b_n-(n+2)\tilde{b}_n\frac{y}{x}]x^{-n-\eta},\\
      c^1(x,y)&=\sum_n c_n x^{-n-1+\eta}, \quad c^0(x,y)=\sum_n [\tilde{c}_n-(n-1)c_n\frac{y}{x}]x^{-n-1+\eta}.
\end{aligned}
\end{equation}
The canonical quantization gives the anti-commutation relation of the modes as
\begin{equation}
    \{b_n,c_m\}=\delta_{n+m,0}, \quad \{\tilde{b}_n, \tilde{c}_m\}=\delta_{n+m,0} \label{anti}
\end{equation}
with all other anti-commutators vanishing.

Now we turn to determine the correlator of $\mathbf{b}$ and $\mathbf{c}$ in the flipped vacuum.  The flipped vacuum  is now defined by
\begin{equation}
\begin{aligned}
     &b_n|0\rangle=\tilde{b}_n|0\rangle=0, \qquad n>-\eta,\\
   &c_n|0\rangle=\tilde{c}_n|0\rangle=0, \qquad n>-1+\eta.  \label{Fvacuum}
\end{aligned}
\end{equation}
Using the anti-commutation relation \eqref{anti}, the vacuum condition \eqref{Fvacuum} and the hermitian relation, we find the following correlators,
\begin{equation}
\begin{aligned}
     &\langle b^0(x_1,y_1)c^1(x_2,y_2)\rangle=0,&\quad \langle b^0(x_1,y_1)c^0(x_2,y_2)\rangle=\frac{1}{x_1-x_2},\\
    &\langle b^1(x_1,y_1)c^1(x_2,y_2)\rangle=\frac{1}{2(x_1-x_2)}, &\quad 
    \langle b^1(x_1,y_1)c^0(x_2,y_2)\rangle=-\frac{y_1-y_2}{(x_1-x_2)^2},  \label{2-ptbc}
\end{aligned}
\end{equation}
and the operator product expansions (OPEs) of $\mathbf{b}$ and $\mathbf{c}$,
\begin{equation}
\begin{aligned}
     &b^0(x_1,y_1)c^1(x_2,y_2)\sim0,
     \quad &b^0(x_1,y_1)c^0(x_2,y_2)\sim \frac{1}{x_1-x_2},  \\ & b^1(x_1,y_1)c^1(x_2,y_2)\sim\frac{1}{2(x_1-x_2)},
     \quad &b^1(x_1,y_1)c^0(x_2,y_2)\sim -\frac{y_1-y_2}{(x_1-x_2)^2}.
\end{aligned}
\end{equation}
Recalling the general form of the 2-point function of a rank-2 multiplet with $\xi=0$,  
\begin{equation}
\begin{aligned}
    &\langle O^0(x_1,y_1)O^0(x_2,y_2)\rangle=0,\\
    &\langle O^0(x_1,y_1)O^1(x_2,y_2)\rangle=\frac{1}{(x_1-x_2)^{2\Delta}},\\
    &\langle O^0(x_1,y_1)O^1(x_2,y_2)\rangle=-\frac{1}{(x_1-x_2)^{2\Delta}}\frac{2(y_1-y_2)}{x_1-x_2},
\end{aligned}
\end{equation}
then we find our result \eqref{2-ptbc} is reasonable, because  
$\Delta_\mathbf{b}+\Delta_\mathbf{c}=1$, which play the role of $2\Delta$ in the above formula for the  2-point function. By Wick's theorem, all the 3-point functions of $\mathbf{b}$ and $\mathbf{c}$ vanish.

Next we study  the $\mathbf{T}\mathbf{b}$ and $\mathbf{T}\mathbf{c}$ OPEs, where $\mathbf{T}=(T,M)^{\mathrm{T}}$ and
\begin{equation}
\begin{aligned}
     T(x,y)&=-:c^0\partial_xb^0-\eta\partial_x(c^0b^0)+2c^1\partial_xb^1-2\eta\partial_x(c^1b^1):\\
     &=:(\eta-1)c^0\partial_x b^0+\eta\partial_x c^0b^0+2(\eta-1)c^1\partial_x b^1+2\eta\partial_x c^1b^1:,\\
     M(x,y)&=-:c^1\partial_x b^0-\eta\partial_x(c^1b^0):\\
     &=:(\eta-1)c^1\partial_x b^0+\eta\partial_x c^1b^0:.
\end{aligned}
\end{equation}
Using the Wick's theorem, we can compute the OPEs with the help of the equations of motion, 
\begin{equation}\nonumber
\begin{aligned}
         M(x_1,y_1)b^0(x_2,y_2)&\sim 0, \\
     M(x_1,y_1)b^1(x_2,y_2)&\sim \frac{\eta b^0(x_2,y_2)}{2(x_1-x_2)^2}+
     \frac{\partial_y b^1(x_2,y_2)}{x_1-x_2},\\
     M(x_1,y_1)c^1(x_2,y_2)&\sim 0,\\
     M(x_1,y_1)c^0(x_2,y_2)&\sim \frac{(1-\eta)c^1(x_2,y_2)}{(x_1-x_2)^2}+\frac{\partial_y c^0(x_2,y_2)}{x_1-x_2},\\
     T(x_1,y_1)b^0(x_2,y_2)&\sim \frac{\eta b^0(x_2,y_2)}{(x_1-x_2)^2}+\frac{\partial_x b^0(x_2,y_2)}{x_1-x_2},\\
    T(x_1,y_1)b^1(x_2,y_2)&\sim \frac{\eta b^1(x_2,y_2)}{(x_1-x_2)^2}+\frac{\partial_x b^1(x_2,y_2)}{x_1-x_2}-\frac{(y_1-y_2)\partial_y b^1(x_2,y_2)}{(x_1-x_2)^2}-\frac{\eta(y_1-y_2) b^0(x_2,y_2)}{(x_1-x_2)^3},\\
      T(x_1,y_1)c^1(x_2,y_2)&\sim \frac{(1-\eta)c^1(x_2,y_2)}{(x_1-x_2)^2}+\frac{\partial_x c^1(x_2,y_2)}{x_1-x_2},\\
     T(x_1,y_1)c^0(x_2,y_2)&\sim \frac{(1-\eta)c^0(x_2,y_2)}{(x_1-x_2)^2}+\frac{\partial_x c^0(x_2,y_2)}{x_1-x_2}-\frac{(y_1-y_2)\partial_y c^0(x_2,y_2)}{(x_1-x_2)^2}-
     \frac{2(1-\eta)(y_1-y_2) c^1(x_2,y_2)}{(x_1-x_2)^3}.
\end{aligned}
\end{equation}
These are precisely the forms of the OPEs of the  stress tensor and a rank-2 multiplet of the quantum number \eqref{general case}. Note that for a general rank-2 multiplet $(O_1,O_0)^{\mathrm{T}}$, there appear additional terms involving  $\partial_yO_0$  in the OPE of $TO_0$ and $MO_0$, here these terms vanish due to the equations of motion.

We can calculate the OPE of the stress tensor with itself directly. Again using the Wick contractions, we read
\begin{equation}\label{OPEofTM}
  \begin{aligned}
    T(x_1,y_1)T(x_2,y_2)\sim& \frac{-3(2\eta-1)^2+1}{(x_1-x_2)^4}+\frac{2T(x_2,y_2)}{(x_1-x_2)^2}+ \frac{\partial_xT(x_2,y_2)}{x_1-x_2}\\
    &-\frac{(y_1-y_2)\partial_yT(x_2,y_2)}{(x_1-x_2)^2}-\frac{4(y_1-y_2)M(x_2,y_2)}{(x_1-x_2)^3},  \\
    T(x_1,y_1)M(x_2,y_2)\sim& \frac{2M(x_2,y_2)}{(x_1-x_2)^2}+\frac{\partial_xM(x_2,y_2)}{x_1-x_2},\\
   M(x_1,y_1)M(x_2,y_2)\sim& 0.
  \end{aligned} 
\end{equation}
This is just the form of the stress tensor OPE in a general BMS field theory. From above, we know that the central charges are:
\begin{equation}
    c_L=-6(2\eta-1)^2+2,\quad \quad c_M=0.
\end{equation}. 

\subsection{Enlarged BMS symmetry}
Apart from the BMS symmetry, there is a ghost number symmetry $\mathscr{G}$, just as in the usual $bc$ ghosts. It is realized as
\begin{equation}
   \mathscr{G}: \quad \delta b^0=-i\epsilon b^0, \quad \delta c^1=i\epsilon c^1, \quad \delta b^1=-i\epsilon b^1, \quad \delta c^0=i\epsilon c^0,
\end{equation}
or in a finite form 
\begin{equation}\label{bcscaling}
   \mathscr{G}_\lambda: \quad  b^0\to  \lambda b^0, \quad  c^1\to \lambda^{-1} c^1, \quad \ b^1\to \lambda b^1, \quad c^0\to \lambda^{-1} c^0.
\end{equation}
The corresponding conserved current is
\begin{equation}\label{bcghostnum}
        J_\mathscr{G}^y=c^0b^0+2c^1b^1,\hspace{3ex}
       J_\mathscr{G}^x=-c^1b^0. 
\end{equation}
Note that the transformation \eqref{bcscaling} is unique in the sense that once the rescaling behaviour of any one of the  ghost fields is determined, the ones of the remaining three fields are determined accordingly. However,  in the usual $bc$ ghosts there are two independent such kind of symmetry transformations: the ghost and anti-ghost number symmetries. Notice the fact that the BMS $bc$ ghosts are the inhomogenous UR limit of the usual $bc$ ghosts, it is expected that there will emerge  another symmetry. This symmetry turns out to be:
\begin{equation}
    \mathscr{G}'_{\theta}: \quad b^0\to b^0, \quad b^1\to b^1-\theta b^0, \quad c^1\to c^1, \quad
    c^0\to c^0+2\theta c^1
\end{equation} 
The corresponding Noether current is:
\begin{equation}
    J_{\mathscr{G}'}^y=-c^1b^0, \qquad  J_{\mathscr{G}'}^x=0.
\end{equation}
Note that $J_{\mathscr{G}'}^y= J_\mathscr{G}^x$. The symmetry $\mathscr{G}_\lambda$ and $\mathscr{G}'_{\theta}$ can be enhanced to local ones:
\begin{equation}\label{local gn}
\begin{aligned}
    \mathscr{G}_f: &\quad  b^0\to  f(x) b^0, \quad  c^1\to f(x)^{-1} c^1, \quad \ b^1\to f(x) b^1+\frac{yf'(x)}{2}b^0, \quad c^0\to f(x)^{-1} c^0-\frac{yf'(x)}{f(x)^2}c^0.\\
    \mathscr{G}'_{g}: &\quad b^0\to b^0, \quad b^1\to b^1-g(x) b^0, \quad c^1\to c^1, \quad
    c^0\to c^0+2g(x) c^1.
\end{aligned}
\end{equation} 
When $f(x)=\lambda$, $g(x)=\theta$, we have $\mathscr{G}_f=\mathscr{G}_\lambda$,  $\mathscr{G}'_{g}=\mathscr{G}'_\theta$. We would like to  comment on the ghost number symmetry here. In the usual string theory, the ghost number symmetry of the $bc$ ghosts could be anomalous when we put them on a general curved background. In the case of the tensionless string, similar anomaly may appear as well. Nevertheless, here we focus on the BMS $bc$ ghost filed theory defined in a flat space-time,  we have not found any anomaly for the ghost number symmetry, similar to the $bc$ ghost in the usual string theory in flat spacetime.

In fact, the symmetry of the BMS $bc$-ghost system is even larger.
In \cite{Yu:2022bcp}, it was shown that in the BMS free scalar and fermion, there is an  anisotropic scaling symmetry. This kind of symmetry also appears in the BMS $bc$-ghost system. It is easy to see that the action \eqref{actionplane} is invariant under the following scaling transformations, which keeps $x$ invariant, 
\begin{equation}\label{scaling xfix}
\begin{aligned}
    \mathcal{D}^\alpha: \qquad &x\to x, \quad y\to \lambda y,\\
    &b^0\to \lambda^\alpha b^0,\quad c^1\to \lambda^{-1-\alpha}c^1,\quad  b^1\to \lambda^{1+\alpha}b^1,\quad c^0\to \lambda^{-\alpha}c^0.
\end{aligned}
\end{equation}
Note that different from the BMS free fermion or scalar, the scaling behaviour of the ghost fields is characterized by a parameter $\alpha$.  As we will see, this is in fact due to the existence of the ghost number symmetry $\mathscr{G}$. One can also find another scaling symmetries $ {\tilde{\mathcal{D}}^\alpha}$, which keeps $y$ invariant, 
\begin{equation}
\begin{aligned}
    \tilde{\mathcal{D}}^\alpha : \qquad &x\to \lambda x, \quad y\to  y,\\
    &b^0\to \lambda^{-\alpha} b^0,\quad c^1\to \lambda^{\alpha}c^1,\quad  b^1\to \lambda^{-1-\alpha}b^1,\quad c^0\to \lambda^{\alpha-1}c^0.
\end{aligned}
\end{equation}
This symmetry is not an independent one because
\begin{equation}
    \tilde{\mathcal{D}}^\alpha=\mathcal{D}\cdot(\mathcal{D}^{-\eta+\alpha})^{-1}, 
\end{equation}
where $\mathcal{D}$ is the isotropic scaling symmetry in the BMS algebra
\begin{equation}\label{isotropic}
\begin{aligned}
\mathcal{D}:\qquad &x\to \lambda x, \quad y\to  \lambda y,\\
&b^0\to \lambda^{-\eta} b^0,\quad c^1\to \lambda^{-1+\eta}c^1,\quad  b^1\to \lambda^{-\eta}b^1,\quad c^0\to \lambda^{-1+\eta}c^0.
\end{aligned}
\end{equation}
So in the following we need only to study $\mathcal{D}^\alpha$. It is easy to derive the corresponding conserved Noether currents of $\mathcal{D}^\alpha$:
\begin{equation}\label{Thecurrent}
\begin{aligned}
    J_{\mathcal{D}^\alpha}^y&=-2c^1\partial_y b^1 y+\alpha c^0b^0+2(1+\alpha)c^1b^1,\\
    J_{\mathcal{D}^\alpha}^x&=-\alpha c^1b^0.
\end{aligned}
\end{equation}


 In fact, 
among all  $\mathcal{D^\alpha}$s,  there is only one independent one, say $D^{\alpha_0}$, and other symmetries are constructed as combinations of $D^{\alpha_0}$ and the ghost number symmetry. To be precise\footnote{We use the subscript to denote the rescaling factor $\lambda$ in \eqref{scaling xfix} and \eqref{bcscaling}.},
\begin{equation}\label{combination}
    \mathcal{D}^\alpha_\lambda =\mathcal{D}^{\alpha_0}_\lambda\mathscr{G}_{\lambda^{\alpha-\alpha_0}}.
\end{equation}
Furthermore, among all the scaling symmetries, there is actually a natural one  with $\alpha_0=-\eta$. It is natural in the sense that only when $\alpha_0=-\eta$,  the OPE among the  current $J^x_{\mathcal{D}^{-\eta}}$ and the stress tensors gives the BMS-Kac-Moody (sub)algebra generated by $L_n, M_n, J^3_n$ in \eqref{full algebra}, which coincide with the space-time BMS-Kac-Moody algebra \eqref{withoutcentral} up to central terms. From \eqref{scaling xfix} and  \eqref{isotropic}, one find that when $\alpha=-\eta$, the dimensions of $b^0$ (or $c^1$) under $ \mathcal{D^\alpha}$ and $\mathcal{D}$  coincide.  Similar phenomenon happens in the BMS free fermion as well\cite{Yu:2022bcp} where the dimensions of $\psi_0$  under the anisotropic scaling and the BMS (isotropic) scaling coincide, both being $\frac{1}{2}$. For later use, we write explicitly the Noether currents of $\mathcal{D}^{-\eta}$:
\begin{equation}\label{anisosym}
\begin{aligned}
    J_{\mathcal{D}^{-\eta}}^y&=-2c^1\partial_y b^1 y-\eta c^0b^0+2(1-\eta)c^1b^1,\\
    J_{\mathcal{D}^{-\eta}}^x&=\eta c^1b^0.
\end{aligned}
\end{equation} 



As in the BMS free scalar and fermion \cite{Yu:2022bcp}, the scaling symmetry $\mathcal{D}^\alpha$ can be enhanced to a local one
\begin{equation}\nonumber
\begin{aligned}
    \mathcal{C}_f^\alpha: \qquad &x\to x, \quad y\to f(x)y,\\
    &b^0\to f^\alpha b^0, \quad c^1\to f^{-1-\alpha}c^1, \quad b^1\to f^{1+\alpha}\left(b^1+\frac{\alpha yf'}{2f}b^0\right),\quad c^0\to f^{-\alpha}\left(c^0-
    \frac{(1+\alpha)yf'}{f}c^1\right).
\end{aligned}
\end{equation}
When $f(x)=\lambda$, we have $\mathcal{C}_f^\alpha=\mathcal{D}^\alpha$.  In fact, these local symmetries have the following decomposition in terms of $\mathcal{C}_f^{-\eta}$ as well as $\mathscr{G}_{f}$ in \eqref{local gn}:
\begin{equation}
    \mathcal{C}_f^\alpha=\mathcal{C}_f^{-\eta}\mathscr{G}_{f^{\eta+\alpha}}
\end{equation} 
Similar to the BMS free fermion \cite{Yu:2022bcp}, the corresponding charges of $\mathcal{C}_f^{-\eta}$ has a  basis of generators\footnote{We use the notation $J^3_n$ to keep consistent with \eqref{modes expan}.} $J_n^3, n\in \mathbb{N}$. They have classical counterparts $j^3_n$ given by the generators of the space-time transformation in $\mathcal{C}_f^\alpha$
\begin{equation}
    j^3_n= -x^ny\partial_y.
\end{equation}
Note that among all $\mathcal{C}_f^{\alpha}$s, only the charges of $\mathcal{C}_f^{-\eta}$ give the quantum version of $j^3_n$, as we have explained below the equation \eqref{combination}.
The generators $j^3_n$, together with the generators of the BMS transformation ($l_n$, $m_n$),
\begin{equation}
\begin{aligned}
    l_n&= -x^{n+1}\partial_x-(n+1)x^{n}y\partial_y\\
    m_n&= -x^{n+1}\partial_y,
\end{aligned}
\end{equation}
form a space-time BMS-Kac-Moody algebra without  central terms\cite{Yu:2022bcp}:
\begin{equation}\label{withoutcentral}
     \begin{aligned}
        &[l_n,l_m]=(n-m)l_{n+m}, \quad &&[l_n,m_m]=(n-m)m_{n+m}, \quad &&[m_n,m_m]=0\\
        &[l_n, j^3_m]=-m j^3_{n+m}, &&[m_n, j^3_m]=-m_{n+m}, &&[j^3_n, j^3_m]=0.\\
    \end{aligned}
\end{equation}
 In the following, we will  determine the central terms in the quantum version of this algebra, by calculating the  OPEs of the corresponding currents.

We have shown that there are three conserved currents which enlarge the BMS symmetry algebra, 
$ J_\mathscr{G}^y$ and $ J_\mathscr{G}^x$ in \eqref{bcghostnum} from the ghost number symmetry $\mathscr{G}$ and $J_{\mathcal{D}^{-\eta}}^y$ in \eqref{anisosym} from the anisotropic scaling symmetry $\mathcal{D}^{-\eta}$. Note that $J_{\mathcal{D}^{-\eta}}^x$ is proportional to $J_\mathscr{G}^x$ so we have in total three independent conserved currents. We denote them as
\begin{equation}
    \mathcal{J}^1\equiv J_\mathscr{G}^y, \quad \mathcal{J}^2\equiv J_\mathscr{G}^x, \quad
    \mathcal{J}^3\equiv J_{\mathcal{D}^{-\eta}}^y.
\end{equation}
  It is obvious that these three currents are of dimension 1, so they must form a  Kac-Moody algebra. In the following, we will work out this Kac-Moody algebra.

\subsection{The full symmetry algebra}
Now we derive the underlying symmetry algebra. In all, there are five currents encoding the symmetries 
\begin{equation}
    T(x,y), \quad M(x), \quad \mathcal{J}^1(x,y), \quad \mathcal{J}^2(x),\quad \mathcal{J}^3(x),
\end{equation}
which have the  modes expansions
\begin{equation}\label{modes expan}
\begin{aligned}
     T(x,y)&=\sum_n L_n x^{-n-2}+\tilde{L}_n x^{-n-3}y,\\
     M(x)&=\sum_n M_n x^{-n-2},\\
    \mathcal{J}^1(x,y)&=\sum_n J^1_n x^{-n-1}+\tilde{J}^1_n x^{-n-2}y,\\
    \mathcal{J}^2(x)&=\sum_n J^2_n x^{-n-1},\\
    \mathcal{J}^3(x)&=\sum_n J^3_n x^{-n-1}+\tilde{J}^3_n x^{-n-2}y.
\end{aligned}
\end{equation}
Note that in the above, the $y$ dependence comes from the equations of motion. As  $\tilde{L}_n=(-n-2)M_n$, $\tilde{J}^1_n=(n+1)J^2_n$, $\tilde{J}^3_n=-\eta(n+1)J^2_n$,  there are in total five kinds of generators in the algebra:
\begin{equation}\label{6generator}
    L_n, \quad M_n, \quad  J^1_n, \quad  J^2_n, \quad  J^3_n.
\end{equation}
From the OPE of $T(x,y)$ and $M(x)$ in \eqref{OPEofTM}, we know that $L_n$'s and $M_n$'s form a BMS algebra, with $c_L=-6(2\eta-1)^2+2, c_M=0$. From the OPE of $\mathcal{J}^1(x,y), \mathcal{J}^2(x), \mathcal{J}^3(x)$, we find that $\{J^1_n,  J^2_n, J^3_n\}$ form a 
 Kac-Moody algebra
\begin{equation}
    [J^a_n, J^b_m]={f^{ab}}_c J^c_n+nk^{ab}\delta_{n+m,0},
\end{equation}
where the structure constant ${f^{ab}}_c$ has only non-vanishing components
\begin{equation}
     {f^{32}}_2=-{f^{23}}_2=1,
\end{equation}
and the level matrix $k^{ab}$ is of the form
\begin{equation}\label{level}
k^{ab}=\left(
   \begin{matrix}
       2 & 0 & 1-2\eta\\
       0 & 0 & 0 \\
       1-2\eta &  0 & \eta^2+(1-\eta)^2
   \end{matrix}\right).
\end{equation}
Note that the matrix of the central terms is degenerate. This is in fact due to the degeneration of the Killing form of the underlying Lie algebra. In other words, this algebra is a non-semisimple Kac-Moody algebra. The underlying Lie algebra is a three-dimensional non-semisimple (non-abelian) Lie algebra generated by  $e^1, e^2, e^3$ with commutation relations
\begin{equation}\label{lie algebra}
    [e^1, e^2]=0, \quad [e^3, e^2]=e^2, \quad [e^1, e^3]=0.
\end{equation}
It has  a maximal abelian ideal generated by $\{e^1, e^2\}$. This algebra is actually  a direct sum  of $\mathbb{R}$ (generated by $e^1$) and a 2-dimensional non-abelian Lie algebra\footnote{Two-dimensional non-abelian Lie algebra is unique and itself is non-semisimple. In fact, the three-dimensional Lie algebra here is of type \uppercase\expandafter{\romannumeral3} in the Bianchi classification of all  three-dimensional Lie algebras.} (generated by $\{e^2, e^3\}$). Remarkably, these two subalgebras no longer decouple when being lifted to the Kac-Moody algebras: $k^{13}$ does not vanish in \eqref{level}.
Notice that the (global) BMS algebra itself is a non-semisimple Lie algebra, thus it is not strange that non-semisimple Kac-Moody algebra appear in BMS field theories.  It is worth pointing out that  a contraction of two  Kac-Moody algebras always leads to non-semisimple Kac-Moody algebras\footnote{ The contraction of two $u(1)$  Kac-Moody algebras  had been studied  in \cite{Bagchi:2022xug}. Non-abelian contractions had been studied recently in \cite{Bagchi:2023dzx}.}. This is similar with the BMS algebra, which is the contraction of two Viraroso algebras and is non-semisimple. Interestingly, the Kac-Moody algebra appear here can $not$ be any contraction of two other Kac-Moody algebras.

 In fact, non-semisimple Kac-Moody algebra appears in the BMS free scalar theory as well. In \cite{Yu:2022bcp}, it was shown that there are two currents with $\Delta=1$: the one associated with the  translational symmetry: $\phi\to\phi+\Lambda(x)$, and the other one  associated with the  anisotropic scaling symmetry,
\begin{equation}
\begin{aligned}
    \mathcal{C}_f:\qquad & x\to x, \quad y\to f(x)y,\\
    &\phi\to f^{\frac{1}{2}}\phi.
\end{aligned}
\end{equation}
The corresponding two currents have the following forms:
\begin{equation}
    j^1=\partial_y\phi, \qquad j^2=\phi\partial_y\phi.
\end{equation}
From the OPE of these currents, it is easy to find the underlying Kac-Moody algebra. It turns out to be:
    \begin{equation}
    [j^a_n, j^b_m]={g^{ab}}_c j^c_n+n\kappa^{ab}\delta_{n+m,0}.
\end{equation}
where the structure constant ${g^{ab}}_c$ has only non-vanishing components
\begin{equation}
     {g^{21}}_1=-{g^{12}}_1=1,
\end{equation}
and the level matrix $\kappa^{ab}$ is of the form
\begin{equation}
k^{ab}=\left(
   \begin{matrix}
        0 & 0 \\
       0 &  -1
   \end{matrix}\right).
\end{equation}
This Kac-Moody algebra is not semisimple and can not be a contraction of any  two other Kac-Moody algebras as well. Its underlying Lie algebra is a two-dimensional subalgebra of \eqref{lie algebra} generated by $(e^2, e^3)$.

Now we turn back to the symmetry algebra of the BMS $bc$ ghosts. We have five sets of generators in \eqref{6generator}: three of them ($ J^1_n,  J^2_n,  J^3_n$) form a Kac-Moody subalgebra, two of them ($L_n, M_n$) form the BMS algebra.   Making use of the  OPE of the five currents, we can  read  the full symmetry algebra:
\begin{equation}\label{full algebra}
    \begin{aligned}
        \left[L_n,L_m\right]&=(n-m)L_{n+m}-\frac{3(2\eta-1)^2-1}{6}n(n^2-1)\delta_{n+m,0},\\
        [L_n,M_m]&=(n-m)M_{n+m},\\
        [M_n,M_m]&=0,\\
         [J^a_n, J^b_m]&={f^{ab}}_c J^c_n+nk^{ab}\delta_{n+m,0},\\
        [L_n, J^1_m]&=-m J^1_{n+m}+(1-2\eta)(n+1)n\delta_{n+m,0},\\
        [L_n, J^2_m]&=-m J^2_{n+m},\\
        [L_n, J^3_m]&=-m J^3_{n+m}+\frac{(2\eta-1)^2}{2}(n+1)n\delta_{n+m,0},\\
        [M_n, J^1_m]&= m J^2_{n+m},\\
         [M_n, J^2_m]&=0,\\
          [M_n, J^3_m]&=-M_{n+m}.
    \end{aligned}
\end{equation}
Note that in the above, the three Kac-Moody generators $J^1, J^2, J^3$ couple differently to the BMS algebra. Notice that there are central terms in $[L_n, J^1_n]$ and $[L_n, J^3_n]$, which means that the currents $J^1$ and $J^3$ are not primaries with respect to the Virasoro subalgebra (except when $\eta=\frac{1}{2}$). This is not strange for a ghost system: recall that in the usual $bc$ ghosts, the ghost number current generally is not a Virasoro primary as well.

Since now the symmetry is enlarged, we want to know the charges of the ghost fields. Besides the BMS charges, the charges of $(b^0, b^1)^{\mathrm{T}}$ with respect to three Kac-Moody currents are
\begin{equation}\label{Jchargeb}
    J_0^1\sim \left(\begin{matrix}
        -1 & 0\\
        0 & -1
    \end{matrix}\right), \qquad 
    J_0^2\sim \left(\begin{matrix}
        0 & 0\\
        \frac{1}{2} & 0
    \end{matrix}\right), \qquad
    J_0^3\sim \left(\begin{matrix}
        \eta & 0\\
        0 & \eta-1
    \end{matrix}\right),
\end{equation}
and the charges of $(c^1, c^0)^{\mathrm{T}}$ are
\begin{equation}\label{Jchargec}
    J_0^1\sim \left(\begin{matrix}
        1 & 0\\
        0 & 1
    \end{matrix}\right), \qquad 
    J_0^2\sim \left(\begin{matrix}
        0 & 0\\
        -1 & 0
    \end{matrix}\right), \qquad
    J_0^3\sim \left(\begin{matrix}
        1-\eta & 0\\
        0 & -\eta
    \end{matrix}\right).
\end{equation}
From above, one can see that  $J^1_0$ is the analog of the usual ghost number. Note that \eqref{Jchargeb} (or \eqref{Jchargec}), together with the BMS charges of $\mathbf{b}$ (or $\mathbf{c}$) in \eqref{general case}, give a two-dimensional representation of the algebra of the zero modes in \eqref{full algebra}.

Finally, we want to comment on the structure of staggered modules. In \cite{Hao:2021urq} and \cite{Yu:2022bcp}, it was shown that  in the BMS free scalar and fermion,  the BMS highest-weight modules could be enlarged to BMS staggered modules or BMS-Kac-Moody staggered modules. It turns out that in the  BMS $bc$-ghost system, the BMS highest-weight modules are enlarged to BMS-Kac-Moody staggered modules, with  \eqref{full algebra} being the underlying BMS-Kac-Moody algebra.   Just like the operator $K(x,y)$ in the  BMS free scalar and fermion, the stress tensor multiplets include the following two operators 
\begin{equation}
    K^1(x,y)\equiv :\partial_x c^0b^1:, \qquad K^2(x,y)\equiv :c^0\partial_xb^1:.
\end{equation}
It is the appearance of these operators that leads to the structure of staggered modules. As the details are similar with the previous two cases, but more cumbersome and  not very instructive, we do not show them here.

\subsection{Inhomogenous BMS \texorpdfstring{$\beta\gamma$}~ ghost}

Another ghost field theory we obtained from the tensionless BMS fermionic string is the BMS $\beta\gamma$ system. The action (on the plane) is:
\begin{equation}
    S=\frac{1}{2\pi}\int dxdy (\gamma_1\partial_0\beta_0-\gamma_0\partial_0\beta_1+\gamma_0\partial_1\beta_0).
\end{equation}
Since the structure of this theory is very similar with the BMS $bc$ ghost system, we will not  give a detailed analysis but  just  give some brief comments on the BMS $\beta\gamma$ ghosts.

It is easy to see that this theory contains two rank-2 boost multiplets: $\boldsymbol{\gamma}=(\gamma_0,\gamma_1)^{\mathrm{T}}$, $\boldsymbol{\beta}=(\beta_0,\beta_1)^{\mathrm{T}}$. They have the following dimensions and boost charges:
\begin{equation} 
\mathbf{\Delta}_{\boldsymbol{\beta}}=\left(\begin{matrix}
   \frac{3}{2} &0  \\
   0& \frac{3}{2}\end{matrix}\right),
  \boldsymbol{\xi}_{\boldsymbol{\beta}}=\left(\begin{matrix}
   0&0  \\
   1&0\end{matrix}\right), \qquad
   \mathbf{\Delta}_{\boldsymbol{\gamma}}=\left(\begin{matrix}
   -\frac{1}{2}&0  \\
   0& -\frac{1}{2}\end{matrix}\right),
  \boldsymbol{\xi}_{\boldsymbol{\gamma}}=\left(\begin{matrix}
   0&0  \\
   1&0\end{matrix}\right).
\end{equation}
One can also study more general BMS $\beta\gamma$ ghosts with dimensions: $\Delta_{\beta_0}=\Delta_{\beta_1}=\chi, \quad \Delta_{\gamma_0}=\Delta_{\gamma_1}=1-\chi$. Then the stress tensors of the generalized theory are
\begin{equation}
\begin{aligned}
      T(x,y)&=(\chi-1)\gamma_0\partial_x\beta_1+\gamma\partial_x\gamma_0\beta_1-(\chi-1)\gamma_1\partial_x\beta_0-\chi\partial_x\gamma_1\beta_0,\\
     M(x,y)&=(\chi-1)\gamma_0\partial_x\beta_0+\chi\partial_x\gamma_0\beta_0.
\end{aligned}
\end{equation}
This theory  has the following central charges:
\begin{equation}
    c_L=6(2\chi-1)^2-2, \qquad c_M=0.
\end{equation}
Similar to the BMS $bc$ ghosts, the BMS $\beta\gamma$ ghosts also has the anisotropic scaling symmetry \cite{Yu:2022bcp}, thus all BMS free theories have this symmetry. It turns out that the full symmetry algebra of the BMS $\beta\gamma$ ghosts is the same  as the one of the  BMS $bc$ ghosts (up to central charges). 

\section{Conclusion and discussion}\label{section5}

In this work, we studied the path-integral quantization of tensionless superstring. We showed how to introduce the BMS $bc$ and $\beta\gamma$ ghosts intrinsically to account for the Faddeev-Popov determinants appeared in the path integrals. For the bosonic string, the BMS $bc$-ghosts can be obtained by taking the inhomogeneous ultra-relativistic limit of the usual $bc$-ghosts. For the tensionless superstring, we found four kinds of $\beta\gamma$-ghosts via intrinsic analysis.  We did canonical quantization of the ghost systems and found the critical dimensions of different tensionless superstrings. It turned out that the critical dimensions of the homogeneous superstring and the doublet inhomogeneous superstring are the same as the usual superstring, but the ones of the trivial homogeneous superstring and the singlet inhomogeneous superstring are not integers. 

We furthermore investigated the BMS ghost systems by taking them as novel kinds of BMS free theories. We showed that the BMS ghosts have enhanced underlying symmetry, just like the BMS free boson and fermion. The underlying symmetry is generated by a BMS-Kac-Moody algebra, with the Kac-Moody subalgebra being built from a three-dimensional non-abelian and non-semi-simple Lie algebra.

  The study in this work, together with the ones on the BMS free scalar and fermion \cite{Hao:2021urq,Yu:2022bcp}, suggests that all BMS free field theories have an enhanced space-time BMS-Kac-Moody symmetry. It would be interesting to see whether this enhancement  persists in  interacting BMS field theories.  Besides, with the BMS $bc$ and $\beta\gamma$ ghosts in hand, it is possible that one can find the bosonization of them.

Before we conclude our work, we would like to discuss the spectrum of tensionless superstring. 
As discussed in detail in \cite{Bagchi:2020fpr}, tensionless bosonic strings have three vacua, which are induced, oscillator and flipped vacuum. The spectrum in the induced vacuum is just the same as the tensile string theory with the tension being taken to zero. In the oscillator vacuum, it is hard to impose the physical conditions. The flipped vacuum has well-defined physical conditions but with the spectrum being truncated. The same things happen in the tensionless superstrings. And the spectrum of homogenous tensionless superstrings has similar structures of tensionless bosonic strings, as shown in \cite{Bagchi:2016yyf,Gamboa:1989zc}. Thus we here focus on the spectrum  of inhomogenous tensionless superstring in the flipped vacuum.

Due to the non-trivial contraction of super-Virasoro algebra in the inhomogenous case, the left-mover and right-mover are coupled with each other, and there does not exist the NS-R or R-NS sector. This feature of inhomogenous theory implies the absence of spacetime supersymmetry. 

In the critical dimension $D=10$, the spectrum of the excitations in the flipped vacuum is truncated at $N_b+\tilde{N}_b+N_f+\Tilde{N}_f=a_L$ where $a_L=1$ for the NS-NS sector and $a_L=0$ for the R-R sector. Here we have defined that
\begin{equation}
\begin{aligned}
    N_b\equiv\sum_{m>0} C_{-m}C_m &,\qquad \tilde{N}_b\equiv-\sum_{m>0} \tilde{C}_{-m}\tilde{C}_m,\\
    N_f\equiv\sum_{r>0} \chi_{-r}\chi_r &,\qquad \tilde{N}_f\equiv-\sum_{r>0} \tilde{\chi}_{-r}\tilde{\chi}_r,\\
\end{aligned}
\end{equation}
with
\begin{equation}
\begin{aligned}
    C_{m}\equiv\frac{1}{2}(X_m+\tilde{X}_m) &,\qquad \tilde{C}_m\equiv\frac{1}{2}(X_m-\tilde{X}_m),\\
    \chi_{r}\equiv\frac{1}{2}(2\psi_r+\tilde{\psi}_r) &,\qquad \tilde{\chi}_r\equiv\frac{1}{2}(2\psi_r-\tilde{\psi}_r).\\
\end{aligned}
\end{equation}
There are two kinds of physical states in the NS-NS sector, 
\begin{equation}
\begin{aligned}
  | \tmop{phy} 1 \rangle \nobracket \nobracket  &=  a_{\mu} C^{\mu}_{- 1} | 0
  \rangle \nobracket \nobracket + a_{\mu}  \tilde{C}^{\mu}_{- 1} | 0 \rangle,
  \nobracket \nobracket\\
  | \tmop{phy} 2 \rangle \nobracket \nobracket & =  h_{\mu \nu} \chi_{-
  \frac{1}{2}}^{\mu} \chi_{- \frac{1}{2}}^{\nu} | 0 \rangle \nobracket
  \nobracket + h_{\mu \nu} \tilde{\chi}_{- \frac{1}{2}}^{\mu} \tilde{\chi}_{-
  \frac{1}{2}}^{\nu} | 0 \rangle \nobracket \nobracket + 2h_{\mu \nu} \chi_{-
  \frac{1}{2}}^{\mu} \tilde{\chi}_{- \frac{1}{2}}^{\nu} | 0 \rangle, \nobracket
  \nobracket
\end{aligned}
\end{equation}
with
\begin{equation}
\begin{aligned}
p^2=0,&\qquad a\cdot p=0,\\
h_{\mu\nu}=-h_{\nu\mu},& \qquad h_{\mu\nu} p^{\nu}=0.
\end{aligned}
\end{equation}
However, all these states are  null, i.e.
\begin{equation}
\langle \nobracket \tmop{phy} 1 | \tmop{phy} 1 \rangle \nobracket \nobracket=\langle \nobracket \tmop{phy} 2 | \tmop{phy} 2 \rangle \nobracket \nobracket=0.
\end{equation}
The physical state in the R-R sector must have $N_{b / f} = \tilde{N}_{b / f} = 0$, so the only physical state is a ``spinor" denoted as $| 0 \rangle_R$, 
transforming under two Clifford algebras, 
\begin{equation}
  \{ \chi_0^{\mu}, \chi_0^{\nu} \}  =  \eta^{\mu \nu},\hspace{3ex}
  \{ \tilde{\chi}_0^{\mu}, \tilde{\chi}_0^{\nu} \}  =  - \eta^{\mu \nu}.
\end{equation}
And the state must satisfy the Dirac-Ramond equation,
\begin{equation}
  p \cdot \chi_0 | 0 \rangle_R \nobracket \nobracket=p \cdot \tilde{\chi}_0 | 0 \rangle_R \nobracket \nobracket  =  0.
\end{equation}

\section*{Acknowledgments}
    We are grateful to  Peng-xiang Hao, Reiko Liu, Haowei Sun for valuable discussions.  The work is supported in part by NSFC Grant  No. 11735001.

\appendix
\renewcommand{\appendixname}{Appendix~\Alph{section}}

\section{Light-cone quantization of inhomogeneous superstring}
In order to crosscheck the critical dimension and the normal ordering constant of the inhomogenous superstring, here we revisit it in the light-cone quantization formalism.
The action of the inhomogeneous superstring is
\begin{equation}
      S  = \frac{1}{2\pi} \int d^2 \sigma [\dot{X}^2 + i (\psi_1 \cdot \dot{\psi_0} + \psi_0
  \cdot \dot{\psi_1} - \psi_0 \cdot \psi_0^{'})],
\end{equation}
which is invariant under local diffeomorphisms and super-transformations
\begin{equation}
    \begin{aligned}
          \delta_{\xi} \psi_0 & =  \xi^a \partial_a \psi_0 + \frac{1}{4} \partial_a
  \xi^a \psi_0,\\
  \delta_{\xi} \psi_1 & =  \xi^a \partial_a \psi_1 + \frac{1}{4} \partial_a
  \xi^a \psi_1 + \frac{1}{2} \partial_1 \xi^0 \psi_0,\\
  \delta_{\xi} X & =  \xi^a \partial_a X,\\
  \delta_{\epsilon} \psi_0 & =  - \epsilon^1 \dot{X},\\
  \delta_{\epsilon} \psi_1 & =  - \epsilon^0 \dot{X} - \epsilon^0 X^{'},\\
  \delta_{\epsilon} X & =  i (\epsilon^0 \psi_0 + \epsilon^1 \psi_1).
    \end{aligned}
\end{equation}
Now we introduce the light-cone coordinates and the worldsheet metric as
\begin{equation}
    \begin{aligned}
          X^{\pm} & \equiv  \frac{1}{\sqrt{2}} (X^0 \pm X^{D - 1}),\\
  \psi^+ & \equiv  \frac{1}{\sqrt{2}} (\psi^0 \pm \psi^{D - 1}),\\
  g_{+ -} & =  - 1,\\
  g_{i j} & =  \delta_{i j}.\\
    \end{aligned}
\end{equation}
Taken the light-cone gauge
\begin{equation}
    \begin{aligned}
  X^+ & =  x^+ + 4 c' p^+ \tau = x^+ + \sqrt{2 c'} B_0^+ \tau,\\
  \psi^+ & =  0,
    \end{aligned}
\end{equation}
then the mode expansions of the rest fields are
\begin{equation}
    \begin{aligned}
  X^i & =  x^i + 4 c' p^i \tau + i \sqrt{2 c^{'}} \sum_{n \neq 0}
  \frac{1}{n} (A_n^i - i n \tau B_n^i) e^{- i n \sigma},\\
  \psi_0^i & =  \sqrt{2 c'} \sum_r \beta_r^i e^{- i r \sigma},\\
  \psi_1^i & =  \sqrt{2 c'} \sum_r (\gamma^i_r - i r \tau \beta_r^i) e^{-
  i r \sigma},
    \end{aligned}
\end{equation}
where $i = 1, 2 \ldots D - 2$. 
The operators appeared in the expansions satisfy canonical quantization commutations 
\begin{equation}
    \begin{aligned}
  {}[x^{\mu}, p^{\nu}] & =  i \eta^{\mu \nu}\\
  {}[A_m^i, B_n^j] & =  2 m \delta_{m + n} \delta^{i j}\\
  \{ \gamma_r^i, \beta_s^j \} & =  2 \delta_{r + s} \delta^{i j},
    \end{aligned}
\end{equation}
and it is useful to define
\begin{equation}
      B_0  \equiv  2 \sqrt{2 c'} p.
\end{equation}
Finally, we need to use the super-BMS constraints to determine the rest fields\footnote{The super-BMS constraints can be derived from the variations of local tetrads and their superpartners.}. The constraints are
\begin{equation}
    \begin{aligned}
  0 = L_n & =  \frac{1}{2} \sum_m A_{- m} \cdot B_{m + n} + \frac{1}{4}
  \sum_r (2 r + n) (\beta_{- r} \cdot \gamma_{r + n} + \gamma_{- r} \cdot
  \beta_{r + n}),\\
  0 = M_n & =  \frac{1}{4} \sum_m B_{- m} \cdot B_{m + n} + \frac{1}{8}
  \sum_r (2 r + n) \beta_{- r} \cdot \beta_{r + n},\\
  0 = G_r & =  \frac{1}{2} \sum_m (A_{- m} \cdot \beta_{m + r} + B_{- m}
  \cdot \gamma_{m + r}),\\
  0 = H_r & =  \frac{1}{2} \sum_m B_{- m} \cdot \beta_{m + r},
    \end{aligned}
\end{equation}
leading to
\begin{equation}
    \begin{aligned}
  A_n^- & =  \frac{1}{2 \sqrt{2 c'} p^+} \sum_m A_{- m}^i B_{m + n}^i +
  \frac{1}{4 \sqrt{2 c'} p^+} \sum_r (2 r + n) (\beta_{- r}^i \gamma_{r +
  n}^i + \gamma_{- r}^i \beta_{r + n}^i)\\
  & \equiv  : A_n^- : - a_L \delta_n,\\
  B_n^- & =  \frac{1}{4 \sqrt{2 c'} p^+} \sum_m B_{- m}^i B_{m + n}^i +
  \frac{1}{8 \sqrt{2 c'} p^+} \sum_r (2 r + n) \beta_{- r}^i \beta_{r +
  n}^i,\\
  \gamma^-_r & =  \frac{1}{2 \sqrt{2 c'} p^+} \sum_m (A_{- m}^i \beta_{m +
  r}^i + B_{- m}^i \gamma_{m + r}^i),\\
  \beta_r^- & =  \frac{1}{2 \sqrt{2 c'} p^+} \sum_m B_{- m}^i \beta_{m +
  r}^i.
    \end{aligned}
\end{equation}
In fact, they form a super-BMS algebra (here we set $c'=\frac{1}{2}$):
\begin{equation}
    \begin{aligned}
  {}[p^+ A^-_m, p^+ A^-_n] & =  (m - n) p^+ A^-_{m + n} + \left[
  \frac{c_L}{12} (m^3 - m) + 2 a_L m \right] \delta_{m + n},\\
  {}[p^+ A^-_m, p^+ B^-_n] & =  (m - n) p^+ B^-_{m + n} + \frac{c_M}{12} (m^3
  - m) \delta_{m + n},\\
  {}[p^+ A^-_m, p^+ \gamma_r^-] & =  \left( \frac{m}{2} - r \right) p^+
  \gamma_{m + r}^-,\\
  {}[p^+ A^-_m, p^+ \beta_r^-] & =  \left( \frac{m}{2} - r \right) p^+
  \beta_{m + r}^-,\\
  {}[p^+ B^-_m, p^+ \gamma_r^-] & =  \left( \frac{m}{2} - r \right) p^+
  \beta_{m + r}^-,\\
  \{ p^+ \gamma_r^-, p^+ \gamma_s^- \} & =  2 p^+ A^-_{r + s} + \left[
  \frac{c_L}{3} \left( r^2 - \frac{1}{4} \right) + 2 a_L \right] \delta_{r +
  s},\\
  \{ p^+ \gamma_r^-, p^+ \beta_s^- \} & =  2 p^+ B^-_{r + s} + \frac{c_M}{3}
  \left( r^2 - \frac{1}{4} \right) \delta_{r + s}.
    \end{aligned}
\end{equation}
In the flipped vacuum, we have
\begin{equation}
    \begin{aligned}
  A_n^i | 0 \rangle & =  B_n^i | 0 \rangle = 0 \quad n > 0,\\
  \gamma_r^i | 0 \rangle & =  \beta_r^i  | 0 \rangle = 0 \quad r > 0,
  \end{aligned}
\end{equation}
and
\begin{equation}
  c_M =  0,\hspace{3ex}
  c_L  =  \frac{3}{2} (D - 2).\end{equation}
    And the mass is determined as
\begin{equation}
    \begin{aligned}
  8 c' M^2 & =  - 8 c' p^2 = 2 B_0^+ B_0^- - B_0^i B_0^i\\
  & =  \sum_{m \neq 0} B_{- m}^i B_m^i + \sum_r r \beta_{- r}^i \beta_r^i,
    \end{aligned}
\end{equation}
where the index "$r$" take value in half-integers in the NS-NS sector and integers in the R-R sector. And here we want to emphasize that there does not exist NS-R sector or R-NS sector due to the non-trivial contraction of the super-Virasoro algebra. Furthermore, this implies that the inhomogeneous superstring does not admit the spacetime supersymmetry anymore and its Green Schwarz formalism seems impossible to construct.

At last, we discuss on the Lorentzian algebra in the inhomogeneous superstring.
Since
\begin{equation}
    \begin{aligned}
  P^{\mu} & \sim  \int \dot{X}^{\mu} d \sigma\sim p^\mu,\\
  J^{\mu \nu} 
  & \sim  \int [\dot{X}^{\mu} X^{\nu} - \dot{X}^{\nu} X^{\mu} + i
  (\psi_1^{\mu} \psi_0^{\nu} - \psi_1^{\nu} \psi_0^{\mu})] d \sigma\\
  & \sim  x^{\mu} p^{\nu} - x^{\nu} p^{\mu}- i \sum_{n > 0} \frac{1}{n} [B_{- n}^{\mu} A_n^{\nu} - B^{\nu}_{- n} A_n^{\mu}]- i \sum_{r > 0} [\beta_{- r}^{\mu} \gamma^{\nu}_r - \beta_{- r}^{\nu}
  \gamma^{\mu}_r].
    \end{aligned}
\end{equation}
Comparing with usual tensile superstring theory, we find that this Lorentzian algebra exactly matches the well-defined UR limit results. This fact implies that the critical dimension and the normal-ordering constant remain the same as usual tensile superstring.

\bibliographystyle{JHEP}
\bibliography{refs.bib}
\end{document}